\documentclass[useAMS,usenatbib,usegraphicx,onecolumn]{mn2e}
\usepackage{enumerate}

\usepackage{subfigure}
\usepackage{color}
\usepackage{float}

\newcommand{\mb}{M_{\bullet}}
\newcommand{\msun}{M_\odot}
\newcommand{\mst}{m_1}
\newcommand{\mnd}{m_2}

\newcommand{\sigmac}{\sigma_{\rm c}}
\newcommand{\kms}{{\rm \, km \, s^{-1}}}
\newcommand{\pc}{{\rm \, pc}}
\makeatletter

\newcommand{\Rmnum}[1]{\expandafter\@slowromancap\romannumeral #1@}
\makeatother

\def\be{\begin{equation}}
\def\ee{\end{equation}}

\def\kms{{\rm \,km\,s^{-1}}}

\def\yr{{\rm \,yr}}

\def\rmb{{\rm b}}

\title[Orbital orientation evolution of massive binary black holes]{Orbital orientation evolution of massive binary black holes at the
centres of non-spherical galaxies
}

\author[X.\ Cui \& Q.\ Yu]{Xiaobo Cui \& Qingjuan Yu\thanks{Author to whom correspondence should
be addressed. Email: yuqj@pku.edu.cn}\\ 
Kavli Institute for Astronomy and Astrophysics, and Department of
Astronomy, Peking University, Beijing, 100871, China }

\begin{document}

\label{firstpage}
\maketitle

\begin{abstract}

At the centre of a spherical and kinematically isotropic galaxy, the
orientation of a massive binary black hole (BBH) orbit (i.e., the direction of
the BBH orbital angular momentum) undergoes a random walk.  If the stars in a
spherical system have a non-zero total angular momentum, the BBH orbital
orientation evolves towards aligning with the total stellar angular momentum
direction. In this paper, we show that a triaxial galaxy has an {\it
alignment-erasing} effect, that is, the alignment of the BBH orientations
towards the galaxy rotation axis can be decreased significantly or erased.  We
also show that in a non-rotating axisymmetric galaxy, the BBH orbital
orientation evolves towards the axisymmetric axis and precesses about it in a
retrograde direction. Our results provide a step towards understanding the spin
orientations of the final merged BH (and hence probable orientation of any jet
produced) within its host galaxy, and may help to constrain the recoiling
velocity of the merged BH arose from gravitational wave radiation as well.

\end{abstract}


\section{Introduction} \label{sec:introduction}

In the modern paradigm of hierarchical galaxy formation and evolution,
formation of massive binary black holes (BBHs) is a natural consequence of
mergers of galaxies with central massive black holes (BHs) (e.g.,
\citealt{BBR1980,yu02}). Study of the BBH orbital evolution and possible
observational signals of existing or existed BBHs is one of the important steps
to understand the formation and evolution of massive BHs, to answer the
possibility of detection of gravitational wave radiation due to the BBH merger
(e.g., \citealt{H09,C10}), and to probe the hierarchical structure formation
model
(e.g., \citealt{VHM03,M07,Y11}). Orbital evolution of BBHs in merged galaxy
remnants has been investigated in various aspects (such as evolution
timescales, evolution of semi-major axes, eccentricities, and orbital
orientations; e.g., \citealt{Q96,yu02,SHM06,sesana10,Merritt02}). In this
paper, we study the orbital orientation evolution of BBHs in purely stellar
systems and investigate how the evolution is related to the BBH host galaxy
properties such as triaxiality and/or rotation.  The orbital orientation is one
of the most basic physical properties of a BBH.  It can influence the spin 
magnitude and 
direction of the merged BH (e.g., \citealt{FH1998}), and the spin direction is
believed to determine the direction of a jet launched from the BH 
inner accretion disk \citep{BP1975, Rees1978}. The BBH orbital orientation can
also influence the gravitational wave radiation from the BBH merger, the
recoiling velocity of the merged BH arose from the asymmetry of the
gravitational wave radiation, and possibly the strength of any electromagnetic
signature of the merger (e.g., \citealt{H09,C10,Bode10,Bode12,B07}).

After a galaxy merger, the orbit of a BBH decays in the merged galaxy remnant.
In a gas-poor environment, the orbital evolution of a BBH can be divided into
several stages, according to the mechanisms that act on the different separation
scales to drain its orbital energy and angular momentum. Initially, each
BH (with mass denoted by $m_1$ or $m_2$, $m_1\ge m_2$) sink independently
towards the
centre of the common gravitational potential under the action of dynamical
friction, at separation scales ranging from several ten kpc to $\sim10$ or
1\pc.  As they migrate inward and form a bound BBH with orbital semimajor axis
\be
a\la r_{\rm inf}\equiv \frac{G \mb}{\sigmac^2} \simeq 10\pc\left( \frac{\mb}{10^8
\, \msun} \right) \left( \frac{\sigmac}{200\kms} \right)^{-2},
\ee
where $\mb=m_1+m_2$ and $\sigmac$ is the one-dimensional velocity dispersion of
the merged galaxy core, it continues to lose energy and angular momentum
through dynamical friction.  However, the influence of dynamical friction on
the BBH orbit becomes less efficient as its orbital period decreases and its
orbital velocity increases.  After the BBH becomes hard at \citep{Q96}
\be
a \sim a_{\rm h}\equiv \frac{G m_2}{4 \sigmac^2} \simeq 2.8 \left( \frac{m_2}{10^8 \,
\msun} \right) \left( \frac{\sigmac}{200 \, \kms} \right)^{-2} \ \pc,
\label{eq:ah}
\ee
it loses energy mainly through three-body interactions with low-angular
momentum stars passing by its vicinity. Finally, after the BBH orbit decays to
some point ($a\la 10^{-2}$ pc), gravitational radiation becomes the dominant
dissipative force (i.e., gravitational radiation stage) to make the BBH lose
energy. The BBH at the different stages has different evolution timescales. The
slowest evolution stage normally starts at $a\sim a_{\rm h}$ and ends at the
gravitational radiation stage, and the evolution timescale at the bottleneck
depends on how many and how fast low-angular momentum stars are available to
interact with the BBH. The non-spherical gravitational potential of galaxies
(e.g., highly flattened or triaxial) has been shown to be effective in having
stars precessing from high-angular momentum orbits onto low-angular momentum
ones so that the bottleneck timescales can be lower than the Hubble time
(\citealt{yu02}; see also \citealt{K13,PBBS11,KJM11,B06}). A BBH is also more
likely to have merged in low-velocity dispersion `power-law' galaxies
\citep{yu02,Z07}.

At the BBH evolution bottleneck, each interaction of the BBH with a star
passing by its vicinity may cause an exchange of the energy and the angular
momentum, and lead to a slight change of the BBH orbital orientation.  The
slight orientation change of each interaction, denoted by $\delta\alpha$, may
accumulate, as the number of the stars passing by (denoted by $N$) increases.
In spherical and isotropic systems, the orbital orientation of a BBH evolves
like a Brownian motion with the accumulated orientation angle change
$\Delta\alpha\propto \sqrt{N\langle\delta\alpha^2\rangle}$
\citep{Merritt02,GM07}. If the stellar system is rotating and the galactic core
has a non-zero total angular momentum, \citet{GDS11} find that the orbital
orientation of a BBH evolves toward the direction of the total angular momentum
of the stars.  Observations reveal that realistic galactic spheroids are likely
to be triaxial (e.g., \citealt{R92,BS00,KY07,Fasanoetal10}). Some recent numerical
simulations (e.g., \citealt{GM12,KJM11,PBBS11}) also show that the stellar
remnant of galaxy mergers is triaxial and rotating.  In this paper, we show
that the non-spherical gravitational potential of a stellar system may affect
the kinematic distribution of the stars passing by the vicinity of the BBH.  We
investigate the orientation evolution of hard BBHs in gas-poor non-spherical
stellar systems including rotating ones, and show how their evolution is
affected by the kinematic distributions of the stars passing by.

The paper is organized as follows. We describe our model and method in
Section~\ref{sec:method}.  We present the simulated orbital orientation
evolution of hard BBHs in different stellar systems (spherical, axisymmetric,
triaxial, rotating) in Section~\ref{sec:results}. As a check for the method, we
apply it first to spherical systems in Section~\ref{subsec:spherical} and find
that the results are consistent with the analytical result and previous work
well. Then we apply the method to non-spherical galaxies.  We present the
kinematic distribution of the stars passing by the vicinity of the BBH in
Section~\ref{subsec:non-spherical}.  By generating a sample of BBHs with
various initial orbital orientations, we illustrate the tendency of the BBH
orientation evolution and distributions after the BBHs interact with a large
number of the stars in Sections~\ref{subsec:BBHevol} and \ref{subsec:BBHdistr}.
The results obtained for non-spherical galaxies are presented along with the
comparison with those obtained for spherical cases. Discussion is given in
Section~\ref{subsec:discussion}, and a summary of the conclusion is in
Section~\ref{sec:summary}.

\section{Model and computational method} \label{sec:method}

In this section, we first present the basic physical equations on the dynamical
evolution of the system to be studied, and then describe the method to solve
the equations. To expedite the calculations, we divide the evolution of the
system into two stages and present them in detail in
Sections~\ref{subsec:precess} and \ref{subsec:scattering}, respectively.

Consider that a massive hard BBH with masses $m_1$ and $m_2$ is located in the
centre of a gas-poor galaxy. We denote the galaxy gravitational potential field
contributed by stars at position ${\bmath r}$ by $\Phi_{\rm G}({\bmath r})$. The BBH
interacts with stars passing by its vicinity through three-body gravitational
interactions, which can be described through the following equations of motion:
\begin{eqnarray} \label{eq:BH1_orbit}
\ddot{\bmath r}_1& = & \frac{G m_2}{|{\bmath r_2} - {\bmath r_1}|^3}({\bmath r_2} - {\bmath r_1})
                +\frac{G m_*}{|{\bmath r_*} - {\bmath r_1}|^3}({\bmath r_*} - {\bmath r_1}), \\
\ddot{\bmath r}_2& = & \frac{G m_1}{|{\bmath r_1} - {\bmath r_2}|^3}({\bmath r_1} - {\bmath r_2})
                +\frac{G m_*}{|{\bmath r_*} - {\bmath r_2}|^3}({\bmath r_*} - {\bmath r_2}), \label{eq:BH2_orbit} \\
\ddot{\bmath r}_*& = & -\bigtriangledown\Phi({\bmath r_*}), \label{eq:star_orbit}
\end{eqnarray}
where ${\bmath r_1}$, ${\bmath r_2}$, and ${\bmath r_*}$ are the position vectors
of the two BHs and the star, respectively, 
\be
\label{eq:potential}
\Phi({\bmath r_*})=- \frac{G \mst}{|{\bmath r_*}-{\bmath r_1}|}
-\frac{G \mnd}{|{\bmath r_*}-{\bmath r_2}|} + \Phi_{\rm G}({\bmath r_*}),
\ee
and the Newtonian mechanics is used.  Note that the galactic potential field
$\Phi_{\rm G}({\bmath r})$ is included in the motion of the star, as the stars
interacting with the hard BBH may come from a large distance in the galaxy and
the stellar motion is affected by the galactic potential along the long ways
to/from the hard BBH. The centre of the galactic potential, i.e., the minimum
point of the potential, is always put at rest at the origin of the coordinate
system in the calculation. The centre of the mass of the BBH is initially put
at the centre of the galactic potential; and the translational Brownian motion
of the BBH is ignored.\footnote{The translational Brownian motion of the BBH is
generally not important in the BBH semi-major axis evolution, although which
might decrease the BBH lifetime in some non-spherical ‘power-law’ galaxies with
low galactic velocity dispersion (\citealt{yu02}; see also \citealt{B05}).  As
seen from this work, the kinematic distribution of the stars interacting
closely with a BBH plays an important role in the BBH orbital orientation
evolution.  As shown in \citet{yu02}, the Brownian wandering amplitudes of the
BBH centres of mass are generally smaller than or comparable to the hardening
radii of the BBHs ($0.01\la q\la 1$); and moreover, the BBHs in high-velocity
dispersion `core' galaxies reach the gravitational radiation stages before
their semimajor axis shrink to the wandering amplitudes of their centres of
mass. Thus the wandering of the BBH should not change the effects of the
stellar kinematic distribution qualitatively on the BBH orbital orientation
qualitatively discussed in our study. Especially, our study shows that the BBH
orbital orientation in triaxial systems can evolve as a Brownian motion,
instead of some alignment along the galactic rotating axis, and the result is
not affected by the BBH wandering through a kinematically isotropic background
induced by the triaxiality of the system (see Section~\ref{sec:results}
below).}

After each interaction of the BBH with a star, the BBH generally receives an
energy loss and angular momentum change, which can be obtained by following
their motion and numerically solving the differential equations
(\ref{eq:BH1_orbit})--(\ref{eq:potential}) together. The total change of the
BBH is an accumulative effect of the interactions. However, the numerical
calculation to follow the stellar motion may be time-consuming, as the moving
time of a star in the galaxy may be much longer than the orbital period of the
BBH and the rotational motion of the BBH described in Equations
(\ref{eq:BH1_orbit}) and (\ref{eq:BH2_orbit}) limits the time steps used in the
calculation. Noting that the galactic potential $\Phi_{\rm G}$ dominates the stellar
motion when the stars are significantly far away from the BBH, the BBH can be
simplified as one object with mass $M_\bullet=m_1+m_2$. Thus, the combined
potential in Equation (\ref{eq:potential}) can be simplified as
\be
\Phi({\bmath r_*})=- \frac{G\mb}{r_*}+\Phi_{\rm G}({\bmath r_*}), \qquad r_*=|{\bmath r_*}|,
\label{eq:potentialsimple}
\ee
which is independent of ${\bmath r_1}$ and ${\bmath r_2}$. To expedite the
calculations, we divide the stellar motion and our calculations into the
following two stages: 
\begin{itemize}
\item stellar precessing stage: the stellar orbits precess in the combined
gravitational potential field described by Equations (\ref{eq:star_orbit}) and
(\ref{eq:potentialsimple}), and we numerically trace the stellar orbits obtain
the kinematic distributions of the stars that can come to a distance $\la a$ to
the galactic centre;
\item three-body scattering stage: we calculate the three-body scattering
processes of the BBH with the stars that can come to its vicinity by using
Equations (\ref{eq:BH1_orbit})--(\ref{eq:potential}). The kinematic
distribution of the stars coming to the BBH vicinity obtained in the above
stellar precessing stage is used as the initial condition of the three-body
interactions.
\end{itemize}

The galactic potential is spherical if it can be described in the form of
$\Phi_{\rm G}(r)$ ($r=|{\bmath r}|$) and triaxial if it can be described in the form
of $\Phi_{\rm G}(x^2+y^2/\xi^2+z^2/\zeta^2)$ ($\zeta<\xi<1$). In our model, we use
the following logarithmic potential as an example for triaxial galaxies:
\begin{equation}
\label{varphi}\Phi_{\rm G} = \sigmac^2 \log(R_{\rm c}^2 + x^2 + \frac{y^2}{\xi^2} + \frac{z^2}{\zeta^2}),
\label{eq:tripotential}
\end{equation}
where $R_{\rm c}$ represents the core radius.
The $\sigmac$ can be set through the tight empirical correlation between
the BH mass and the galactic velocity dispersion (e.g., \citealt{T02,FR05,G09,MM13}),
and in this paper we adopted the following relation:
\begin{equation}
M_{\bullet}=1.66\times10^8\msun \left(\frac{\sigmac}{200\kms}\right)^{4.86}.
\label{eq:msigma}
\end{equation}
Although a self-consistent model of a realistic stellar distribution around a
massive BH is not used in this paper, the above logarithmic potential used
should be sufficient to display the effects of the triaxiality qualitatively.

\subsection{Stellar precessing stage}\label{subsec:precess}

We describe the dynamics of a star in the phase space of its specific energy
and specific angular momentum $(E,J)$.  In spherical potentials, the angular
momenta of stars are conserved, and the stars that can come to the vicinity of
the BBH ($\sim a$ from the centre) have the orbital angular momenta $J<
J_{\rm lc}$, where $J_{\rm lc}\simeq \sqrt{2G\mb a}$ and the corresponding region in
the phase space is called the ``loss cone''.  In axisymmetric and triaxial
galaxies, there exist centrophilic orbits such as box orbits, which pass
arbitrarily close to the centre and have low angular momentum, as well as
centrophobic orbits such as loop orbits, which avoid centre and have high
angular momentum.  Here, we introduce $J_{\rm s}$ to mark the transition from
centrophilic ($J\la J_{\rm s}$) to centrophobic ($J\ga J_{\rm s}$) orbits.  Stars on
centrophilic orbits with $J<J_{\rm s}$ can process into the loss cone. In
non-spherical galaxies, the number of stars available to interact with the BBH
can be large, and it has been shown that the BBH bottleneck evolution timescale
can be decreased significantly (\citealt{yu02}; see also
\citealt{PBBS11,KJM11}).

We use Monte-Carlo simulations to obtain the characteristic angular momentum
$J_{\rm s}$ and the kinematic distribution of the stars that can come to the vicinity
of the BBH. In our simulations, the initial kinematic settings for the stars
are set as follows.
\begin{itemize}
\item Given the specific energy $E$ of a star, its specific angular momentum
$J$ is randomly generated so that $J^2$ is uniformly distributed within the
range $[0, J_{\rm c}^2]$, where $J_{\rm c}$ is the specific angular momentum of a circular
orbit at energy $E$ (cf., Eq.\ 4.288 in \citealt{BT08}, where the difference in
stellar orbital periods with different $J$ is ignored).
\item The test particle is initially put at its apocentre with zero radial
velocity. Given its $(E,J)$, its apocentre distance to the central BH $r$ and
velocity $v$ can be determined by approximating the potential as spherical
through setting $\xi=\zeta=1$ in the triaxial potential.
The relative error of the initial $r$ and $v$ caused by approximating the
potential as spherical is negligible for the purpose of this paper, as the
orbital energy difference for the same initial $r$ and $v$ between the
spherical and the triaxial/axisymmetric systems is at the most the galactic
potential difference between the positions ${\bmath r_*}=(0,0,r)$ and
$(0,0,r/\zeta)$. The initial direction
of the orbital angular momentum of the star is parameterized by $(\theta_*,
\phi_*)$, where $\theta_*$ is the angle between the angular momentum and the
$z$-axis and $\phi_*$ is the azimuthal angle of the angular momentum direction
in the $x-y$ plane.  The directions are generated isotropically and randomly.
The argument of its apocentre $\psi_*$ is generated randomly within $[0,2\pi]$.
Thus, the position and velocity vectors of stars can be derived and serve as
the initial conditions for Equation (\ref{eq:star_orbit}).
\end{itemize}
We use the above initial settings and Equations (\ref{eq:star_orbit}) and
(\ref{eq:potentialsimple}) to trace the motion of the stars. The calculation
for each star is terminated once its $r\la a$ (centrophilic orbits) or its
traveling time is longer than $2\times 10^5 GM_\bullet/\sigmac^3\simeq
1.4\times10^{10}\yr(M_\bullet/10^8M_\odot)^{0.38}$ (where Eq.~\ref{eq:msigma}
is used). We select those stars for which the calculation is terminated due to
$r\la a$, and obtain their kinematic distributions when they pass through
$r\simeq r_{\rm inf}$ at its last orbit of the calculation. The kinematic
distributions of these stars are used as the initial conditions in the
calculation for the three-body scattering stage of the stars interacting with
the BBH.  The $J_{\rm s}$ can be determined from the initial angular momentum
distribution of the stars (see Fig.~\ref{fig:Jstriaxial} later).

\subsection{Three-body scattering stage}\label{subsec:scattering}

Given a BBH with total mass $M_{\bullet}$, mass ratio $q\equiv m_2/m_1$, we set
its semi-major axis $a=a_{\rm h}$. The unit orbital angular momentum of the BBH is
denoted by $\hat{\bmath l}_\rmb=(l_{\rmb,x},l_{\rmb,y},l_{\rmb,z})$, and the orientation
direction can also be parameterized by angles $(\theta_\rmb,\phi_\rmb)$, similarly as
the angles $(\theta_*,\phi_*)$ described above for the stellar angular
momentum.  We randomly select a star from the kinematic distribution obtained
in the stellar precessing stage and use Equations
(\ref{eq:BH1_orbit})--(\ref{eq:star_orbit}) to trace its three-body scattering
processes with the BBH. The calculation is terminated when the distance of the
star has $r\ga 2r_{\rm inf}\simeq 8(1+q^{-1})a_{\rm h}$ or its traveling time in the
three-body scattering stage is longer than $10 GM_\bullet/\sigmac^3$, and the
BBH orbital orientation $(\theta_\rmb, \phi_\rmb)$ is recorded and used as the
initial condition in the scattering process with next star. The calculation is
reiterated by scattering with $N$ number of stars. The number of the scattered
stars is related to the BBH orbital decay by the following equation (see Eq.\
12 in \citealt{yu02}):
\be
N\simeq 0.32\frac{\mb}{m_*}\ln\left(\frac{a_{\rm h}}{a}\right).
\label{eq:nscatter} \ee
For simplicity and saving the calculation time, the semi-major axis and the
eccentricity of the BBH $(a,e)$ is fixed during the reiterations; and the
translational Brownian motion of the BBH induced after the scattering process
with each star is ignored, with the center of mass of the BBH being reset to the
centre of the galactic potential at the beginning of the scattering with the
next star. We also assume that the stars have an identical stellar
mass $m_*$(=$10^{-4}M_\bullet$). All of the simplifications made above do not
affect our main conclusions (see discussion in
Section~\ref{subsec:discussion}).  

Regarding the conditions on the traveling time that are set to terminate the
numerical calculations in the above two stages, we have tested that the results
below are not affected much if the time lengths are increased by one order of
magnitude. With these settings in natural units of the dynamical system, the
results below do not depend on the detailed values of $M_\bullet$ 

In the calculations, the motion of the stars and the BBH is traced by
integrating the differential equations of their motion with an explicit
Runge-Kutta method of order 8(5,3) \citep{H93,DP78}.
In the three-body scattering stage, the total energy of the BBH and the star is
conserved; and to ensure the accuracy of numerical calculations, the change in
their total energy due to numerical errors must be much smaller than the change
of the BBH energy due to each interaction with a star passing by, so that the
change of the BBH energy obtained from the calculations is not caused by
numerical errors.  The absolute relative error of the total energy achieved in our
calculations is lower than $\sim 10^{-9}$ for each three-body encounter, which
is accurate enough, as the relative change in the BBH energy due to each
interaction with a passing-by star is roughly order of $m_*/m_1$.  We have also
checked that our results are not affected much even by decreasing the accuracy
by one order of magnitude. As to be mentioned in Section~\ref{sec:results}, the
validity of our calculation methods is further supported by
Figure~\ref{fig:deltathetaspherical} below.

\section{Results} \label{sec:results}

\subsection{Simple model test: spherical galaxies}\label{subsec:spherical}

We apply the model and the numerical method described in
Section~\ref{sec:method} to spherical galaxies. The obtained BBH orientation
evolution serves as our model test, and we find that they are consistent with
theoretical expectation well. 

\begin{figure} \begin{center}
\subfigure[]{\includegraphics[width=0.4\textwidth]{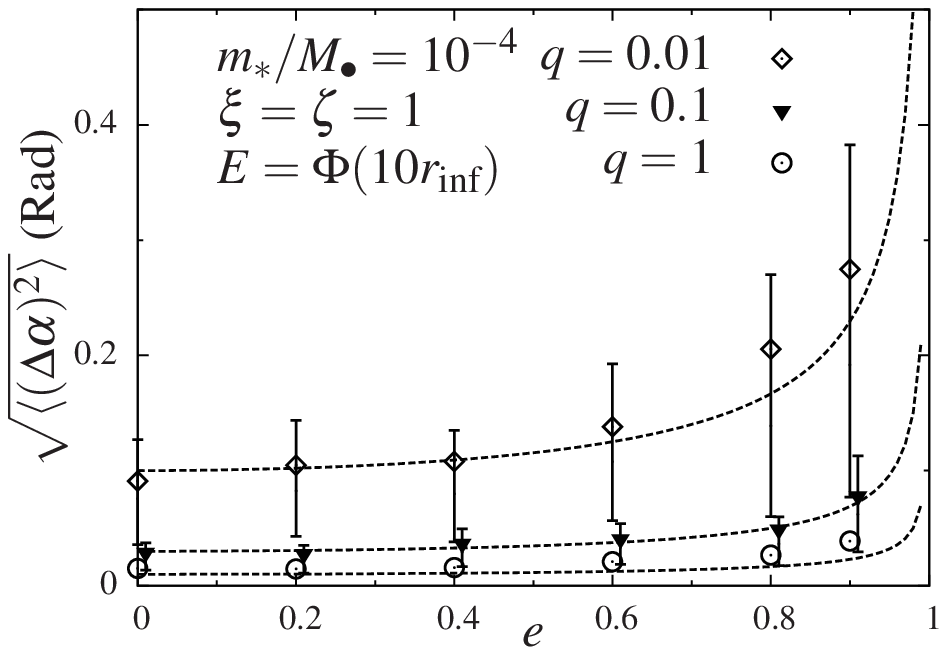}}
\subfigure[]{\includegraphics[width=0.4\textwidth]{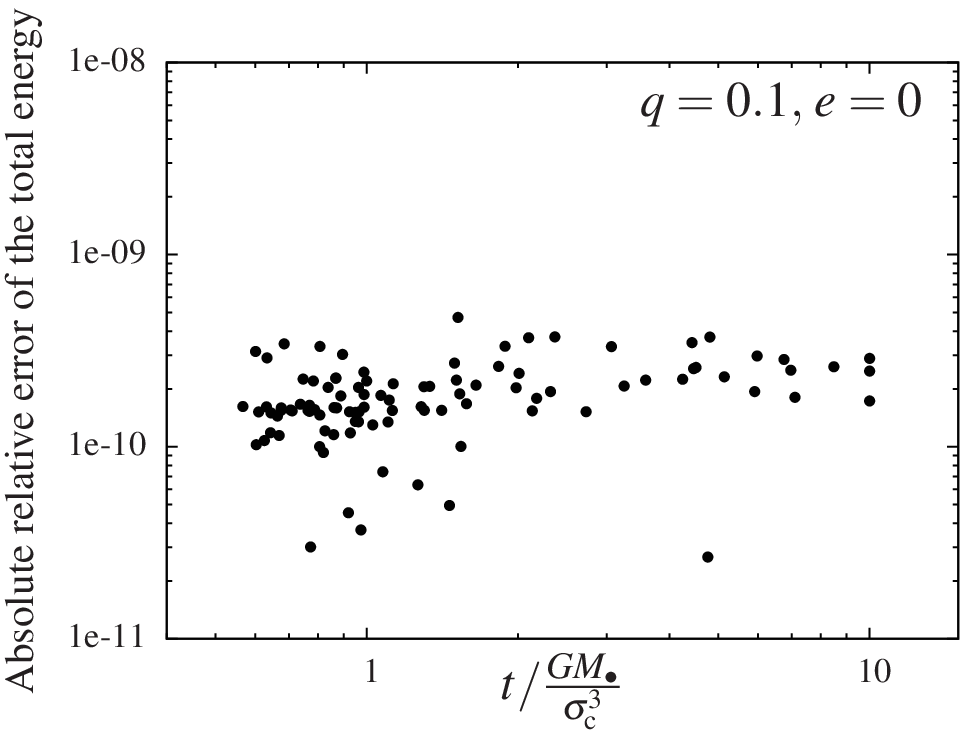}}
\end{center}
\caption{Panel (a): the rms of the orientation change of hard BBHs in spherical
systems in one hardening time. The points are our simulation results with
different BBH mass ratio $q(=0.01,0.1,1)$ and eccentricity
$e(=0,0.2,0.4,0.6,0.8,0.9)$. Each point is the rms of orientation change of 100
hard BBHs with initial orientations generated randomly,
$\langle(\Delta\alpha)^2\rangle^{1/2}$, and the error-bar of each point
represent the range bounded by the lowest 15\% and the highest 15\% values of
the 100 orientation changes.  For view clarity, the points for $q=0.1$ are
shifted rightwards a little, and the error-bars only for $q=0.01,0.1$ are shown
in the figure. The dashed curves are the analytical expectation from Equation
(\ref{eq:delta_theta}).  As seen from this panel, our simulation results are
well consistent with Equation (\ref{eq:delta_theta}), although it shows a
relatively large deviation at small $q$ and high $e$. Panel (b): an example for
the absolute relative error of the total energy achieved for each three-body
scattering in our calculation, which is lower than $\sim 10^{-9}$. The
horizontal axis represents the traveling time of each star at the three-body
scattering stage, in unit of $GM_\bullet/\sigmac^3$. For simplicity, we 
only show the errors of 100 simulation results randomly chosen from the case
of $q=0.1$ and $e=0$.
This panel serves as a supplementary support for the high accuracy in our
calculation. }
\label{fig:deltathetaspherical} \end{figure}

In our calculations for the example case, the related parameters are set as
follows: $\xi=\zeta=1$, $R_{\rm c}=4 r_{\rm inf}$, the BBH mass ratio $q\equiv
m_2/m_1=0.01,0.1,1$, $m_*=10^{-4}M_\bullet$. The semi-major axis of the BBH is
chosen to be $a_{\rm h}$, and the eccentricity $e$ is set to $0, 0.1, \cdots, 0.9$.
All the stars are set to have the same specific energy (e.g.,
$E=\Phi(10r_{\rm inf})$ here), and the results are not sensitive to the detailed
value of $E$ if $E$ is high enough. In one hardening time
$t_{\rm h}\equiv|a/(da/dt)|$, the number of the scattered stars is $N\simeq
0.32(M_\bullet/m_*)\simeq 3200$\footnote{Note that although a full mapping of
the orbits in a logarithmic potential needs a large number of orbits, the
several thousand orbits used here should be sufficient to map the orbits at the
given energy that can pass by the vicinity of the BBH. Even if there exist
some delicate orbits which are not contained in these several thousand orbits,
the $10^4$ orbits used in
Figs.~\ref{fig:precessiontriaxial}--\ref{fig:precessiontrirotation}, or the
$2\times10^4$ orbits used in
Figs.~\ref{fig:BBHevolspherical}--\ref{fig:BBHevolnonspherical}, they should
not affect the main results significantly and statistically.} (see
Eq.~\ref{eq:nscatter}).  We simulate the orientation changes of the BBH within
one hardening time. For each set of the BBH mass ratio and eccentricity, we do
the Monte-Carlo simulation for 100 times and show the root mean square (rms) of
the orientation change in Figure~\ref{fig:deltathetaspherical}.  As seen from
Figure~\ref{fig:deltathetaspherical}, our calculated BBH orientation changes
and their dependence on $q$ and $e$ (especially at low $e$) is generally well
consistent with the following analytical expectation \citep{GM07}:
\be \label{eq:delta_theta}
\Delta \alpha \sim q^{-1/2} \left( \frac{m_\ast}{M_\bullet} \right)^{1/2}
(1-e^2)^{-1/2},
\ee
although they have a relatively large deviation at small $q$ and high $e$. The
relatively large deviation are unlikely to be caused by numerical errors in the
calculations, as they do not change much by changing the numerical accuracy by
one order of magnitude; and they are more likely to be due to some
approximation in deriving the analytical formula. We find that the values of
$\Delta\alpha$ of the 100 BBHs scatters largely at small $q$ and high $e$, and
the medians of the $\Delta\alpha$ distributions are consistent with Equation
(\ref{eq:delta_theta}) better.

For convenience, the parameter sets used in
Figure~\ref{fig:deltathetaspherical} and some other figures below are listed in
Table~\ref{tab:para}.

\begin{table*}
\begin{tabular}{lcccccccc} \hline \hline
Figure   & Shape        & Rotation & $q$        & $e$   & $E$                       & $N$   & BBH orient.\\ \hline
1        & spherical    & N        & 1,0.1,0.01 & 0-0.9 & $\Phi(10 r_\mathrm{inf})$ & 3200  & 100 \\ \hline \hline
2        & triaxial     & N        & 0.1        &   N/A & $\Phi(2 r_\mathrm{inf})$  & 10000 & N/A\\
         &              &          &            &       & $\Phi(5 r_\mathrm{inf})$  &       &    \\
         &              &          &            &       & $\Phi(10 r_\mathrm{inf})$ & & \\ \hline 
3(top)   & triaxial     & N        & 0.1        &   N/A & $\Phi(2 r_\mathrm{inf})$  & 10000 & N/A\\
3(middle)&              &          &            &       & $\Phi(3 r_\mathrm{inf})$  &       &    \\
3(bottom)&              &          &            &       & $\Phi(10 r_\mathrm{inf})$ & & \\ \hline
4(top)   & axisymmetric & N        & 0.1        &   N/A & $\Phi(2 r_\mathrm{inf})$  & 10000 & N/A  \\
4(middle)&              &          &            &       & $\Phi(3 r_\mathrm{inf})$  & & \\
4(bottom)&              &          &            &       & $\Phi(10 r_\mathrm{inf})$ & & \\ \hline
5(top)   & triaxial     & Y        & 0.1        &   N/A & $\Phi(2 r_\mathrm{inf})$  & 10000 & N/A  \\
5(middle)&              &          &            &       & $\Phi(3r_\mathrm{inf})$   &  & \\
5(bottom)&              &          &            &       & $\Phi(10r_\mathrm{inf})$  &  & \\ \hline \hline
6(a)     & spherical    & N        & 0.1        &   0   & $\Phi(10 r_\mathrm{inf})$ & 20000 & 7  \\
6(b)     &              & Y        &            &       & &        &  \\ \hline %
7(a)     & triaxial     & N        & 0.1        &   0   & $\Phi(3 r_\mathrm{inf})$  & 20000 & 7 \\
7(b)     & triaxial     & Y        &            &       & $\Phi(3 r_\mathrm{inf})$  &       & \\
7(c)     & triaxial     & Y        &            &       & $\Phi(10 r_\mathrm{inf})$ & & \\
7(d)     & axisymmetric & N        &            &       & $\Phi(10 r_\mathrm{inf})$ & & \\\hline 
8(a)     & spherical    & Y        & 0.1        &   0   & $\Phi(10 r_\mathrm{inf})$ & 20000 & 1500 \\
8(b)     & triaxial     & N        &            &       & &   &  \\
8(c)     & axisymmetric & N        &            &       & &   &  \\\hline \hline
9        & spherical    & Y, N     & 0.1        &   0 & $\Phi(10 r_\mathrm{inf})$ & 3200  & 100 \\ 
         & axisymmetric & N        &            &     &                           &       &     \\ \hline
\end{tabular}
\medskip
\caption{Parameter sets used in some figures. The first column gives the figure
number and the corresponding panel (if any). The second column gives the shape
of the galactic gravitational potential used: `spherical' corresponds to
$(\xi,\zeta)=(1,1)$, `triaxial' corresponds to $(\xi,\zeta)=(0.9,0.8)$ in
Equation (\ref{eq:tripotential}), and `axisymmetric' corresponds to
$(\xi,\zeta)=(1,0.8)$. The third column represents the rotational property set
to the galaxy, where the labels `Y' and `N' mean a rotating galaxy ($P_+=7/8$)
and a non-rotating one ($P_+=1/2$), respectively. The $q$ and $e$ are the BBH
mass ratio and eccentricity.  For
Figs.~\ref{fig:Jstriaxial}--\ref{fig:precessiontrirotation}, the BBH eccentricity
is not needed to obtain the kinematic distributions of the stars that can
precess into the loss cone during the stellar precession stage, so it is
labeled by `N/A' (similarly for the last column), but the value of $q$ is
needed to define the size of the loss cone by giving the semimajor axis of a
hard BBH. The $E$ represents the specific energy of the stars used in the
simulation. Given the shape and rotational property of a galaxy, the $N$
represents the number of the scattered stars used given each parameter set of
the stellar energy and the BBH configuration in the simulations of each figure,
where the scattered stars are those that can pass by the vicinity of the BBH.
The last column represents the number of the BBH orientations randomly
generated given all the other parameters; and in
Figs.~\ref{fig:BBHevolspherical}--\ref{fig:BBHevolnonspherical}, only seven cases are
shown by being randomly selected from 100 simulations.  In all the
figures, we set $m_*=10^{-4}M_\bullet$ and $R_{\rm c}=4r_{\rm inf}$. The blank in the
table means that the parameter is the same as that shown in the first row of
the corresponding figure.  } \label{tab:para} \end{table*}

\subsection{Dynamical distribution of stars that can precess onto the loss
cone in non-spherical galaxies} \label{subsec:non-spherical}

\subsubsection{Triaxial and axisymmetric galaxies} \label{subsec:dyndis}
 
\begin{figure}
\includegraphics[width=0.8\textwidth]{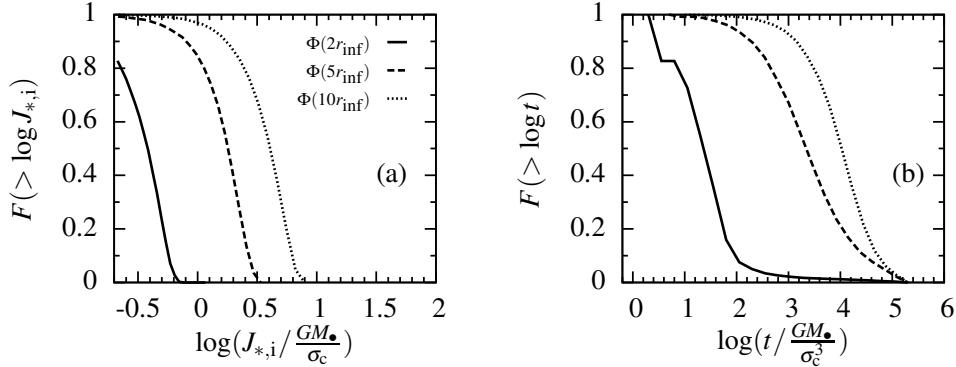}\centering
\caption{ (a) Example for the cumulative distribution of the initial angular
momenta of the stars that can precess to the vicinity of a BBH in a triaxial
galaxy, with parameters $q=0.1$, $R_{\rm c}=4r_{\rm inf}$, and
$(\xi,\zeta)=(0.9,0.8)$.  The angular momentum is in unit of
$GM_\bullet/\sigmac$. The lower boundary of the $x$-axis marks the loss cone of
the BBH at its hardening radius in a spherical system, i.e., $\log(J_{\rm
lc}/\frac{GM_\bullet}{\sigma_{\rm c}})\simeq
\log\sqrt{\frac{m_2}{2M_\bullet}}\simeq -0.7$. (b) The cumulative distribution
of the traveling time for the stars to precess to the vicinity of the BBH, in
unit of $GM_\bullet/\sigmac^3$.  Different curves represent the stars with
different specific energy. The number of the simulated stars used for each
curve is 10000. The Hubble timescale is located at
$5.2-0.38\log(M_\bullet/10^8\msun)$ in the $x$-axis (where Eq.~\ref{eq:msigma}
is used). The figure indicates that the number of the stars in a triaxial system
that can move into the vicinity of a BBH within a Hubble time can be much more
than those in spherical systems, and thus the triaxiality of the stellar system
can play an important role in shrinking the BBH orbit. See details in Section~\ref{subsec:dyndis}.  }
\label{fig:Jstriaxial}
\end{figure}

\begin{figure}
\includegraphics[width=0.8\textwidth]{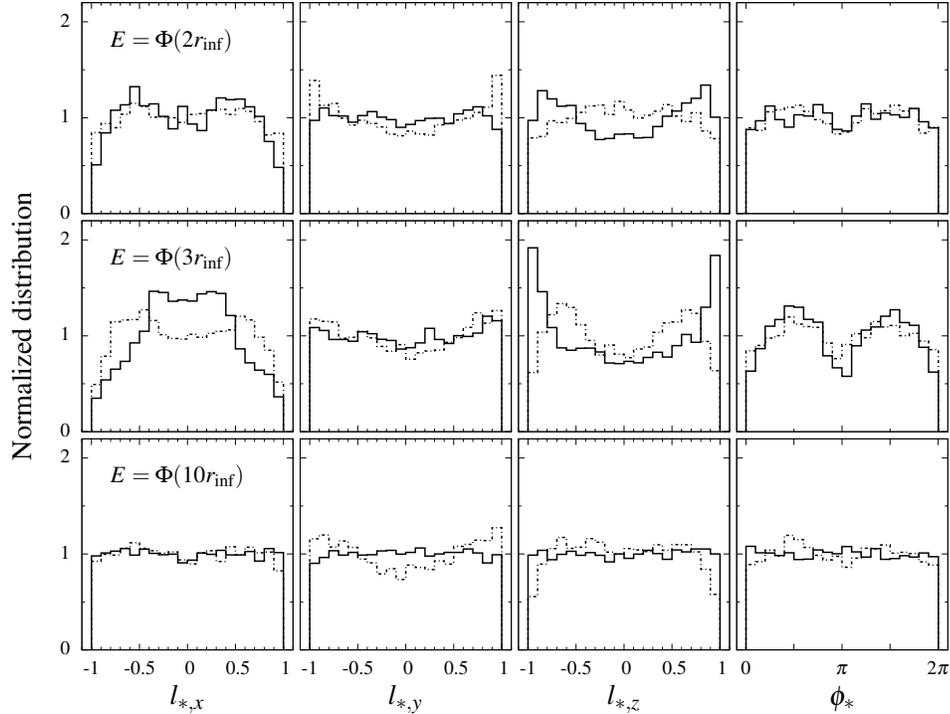}\centering
\caption{Histograms of the angular momentum orientation distribution of the
stars that can precess into the loss cone during the stellar precession stage
in triaxial galaxies. The vertical axis represent the number fraction of the
stars normalized by the average fraction in each horizontal axis bin.  The
parameters $(q,R_{\rm c},\xi,\zeta)$ are the same as those in
Figure~\ref{fig:Jstriaxial}.  The number of the stars used for the statistics
in each row is 10000. The dotted lines represent the initial distributions and
the solid lines represent the final distributions. Different rows show the
results for the stars with different specific energies. As seen from the
figure, in triaxial systems the final kinematic distribution at the stellar
precessing stage can be isotropic for high-energy stars (bottom panel) and
anisotropic for low or intermediate-energy stars (top or middle panel).
} \label{fig:precessiontriaxial}
\end{figure}

We use the model and the numerical method described in Section~\ref{sec:method}
to obtain the kinematic distribution of the stars that can precess onto the
loss cone in triaxial galaxies. We illustrate our calculation results in
Figures~\ref{fig:Jstriaxial} and \ref{fig:precessiontriaxial}, where $q=0.1$,
$R_{\rm c}=4r_{\rm inf}$, and $(\xi,\zeta)=(0.9,0.8)$. The values of
$(\xi,\zeta)$ fall well into the observational distribution of the axis ratios
of elliptical galaxies and brightest cluster galaxies (e.g.,
\citealt{Fasanoetal10,KY07,R92}), where we note that the axis ratios used here
are for the isopotential shape, not for the intrinsic isophotal shape of a
galaxy, and the flattening in the potential is roughly a third of that in the
density distribution (see eq.~2.72b in \citealt{BT08}).

Figure~\ref{fig:Jstriaxial}(a) shows the cumulative distribution of the initial
specific angular momenta of the stars, where different curves represent
different initial energy of the stars. In the panel, the drop-off at the
high-angular momentum end characterizes the transition of centrophobic orbits
to centrophilic orbits, as $J_{\rm s}$ mentioned in Section~\ref{sec:introduction}.
In spherical systems, we have $J_{\rm s}\simeq J_{\rm lc}$, which is about
$0.2GM_\bullet/\sigmac$ for the example shown in Figure~\ref{fig:Jstriaxial}.
In triaxial systems, $J_{\rm s}$ can be much larger than $J_{\rm lc}$, especially for
stars with high energy (or large apocentre distances) whose motion is affected
more significantly by the triaxiality of the potential.
Figure~\ref{fig:Jstriaxial}(b) shows the cumulative distribution of the
traveling time taken for the stars to precess into the loss cone.  The
traveling time of a significant fraction of stars is less than several thousand
times of $G\mb/\sigmac^3$, generally shorter than the Hubble time.

For the stars shown in Figure~\ref{fig:Jstriaxial}, we show the initial and the
final distributions of their orbital orientations reached at the stellar
precessing stage in Figure~\ref{fig:precessiontriaxial} (see
Section~\ref{sec:method}). The stellar orbital orientations are expressed
through the unit vector of their angular momenta $(l_{*,x},l_{*,y},l_{*,z})$,
and the angle $\phi_*$ is defined as the azimuthal angle of the vector
projected onto the $x$-$y$ plane. The distributions are shown for different
specific energies of the stars. In spherical distributions, the final
distributions should be isotropic if the initial distributions are isotropic,
due to the conservation of the angular momentum; and the distribution curves
should be flat.  However, in triaxial systems, as seen from the figure, the
curves in many cases are not flat and the anisotropy of stellar distributions
are displayed in the following two aspects. First, the probability for stars
with different initial orbital orientations to precess into the loss cone is
not identical (see dotted line).  Second, the distribution of the orbital
orientation of the stars when they move to the vicinity of the central BHs is
anisotropic (see solid line).  For the low energy case ($E\simeq
\Phi(2r_{\rm inf})$; top panel), although this anisotropy appears mild, it does
exist.  For relatively high energy (middle panel), the anisotropy appears more
obvious, and a significant fraction of stellar orbits orient close to the short
axis ($z$-axis).  The final distributions deviate significantly from the
initial ones, which manifests the effect of the torque induced from the
triaxial potential.  However, for even higher energies or larger apocentre
distances ($E \ga \Phi(10r_{\rm inf})$; bottom panel), the distribution curves are
close to be flat and the anisotropy decreases significantly.  Note that the
centrophilic orbits with relatively low energy are regular and those with
relatively high energy are stochastic; and the critical energy for the
transition of the orbits depends on the detailed shape of the gravitational
potential (e.g., the parameters $(R_{\rm c},\xi,\zeta)$ for the model used here).
The degree of the anisotropy of the stellar kinematic distribution is related
with the relative fraction of regular centrophilic orbits and stochastic ones.
The angular momenta of regular orbits have their own regular precession
patterns and are more likely to indicate the anisotropy in the distribution, in
contrast to stochastic orbits.  The anisotropy disappears at the high energy
because of the dominance of the stochastic orbits. 

\begin{figure}
\includegraphics[width=0.8\textwidth]{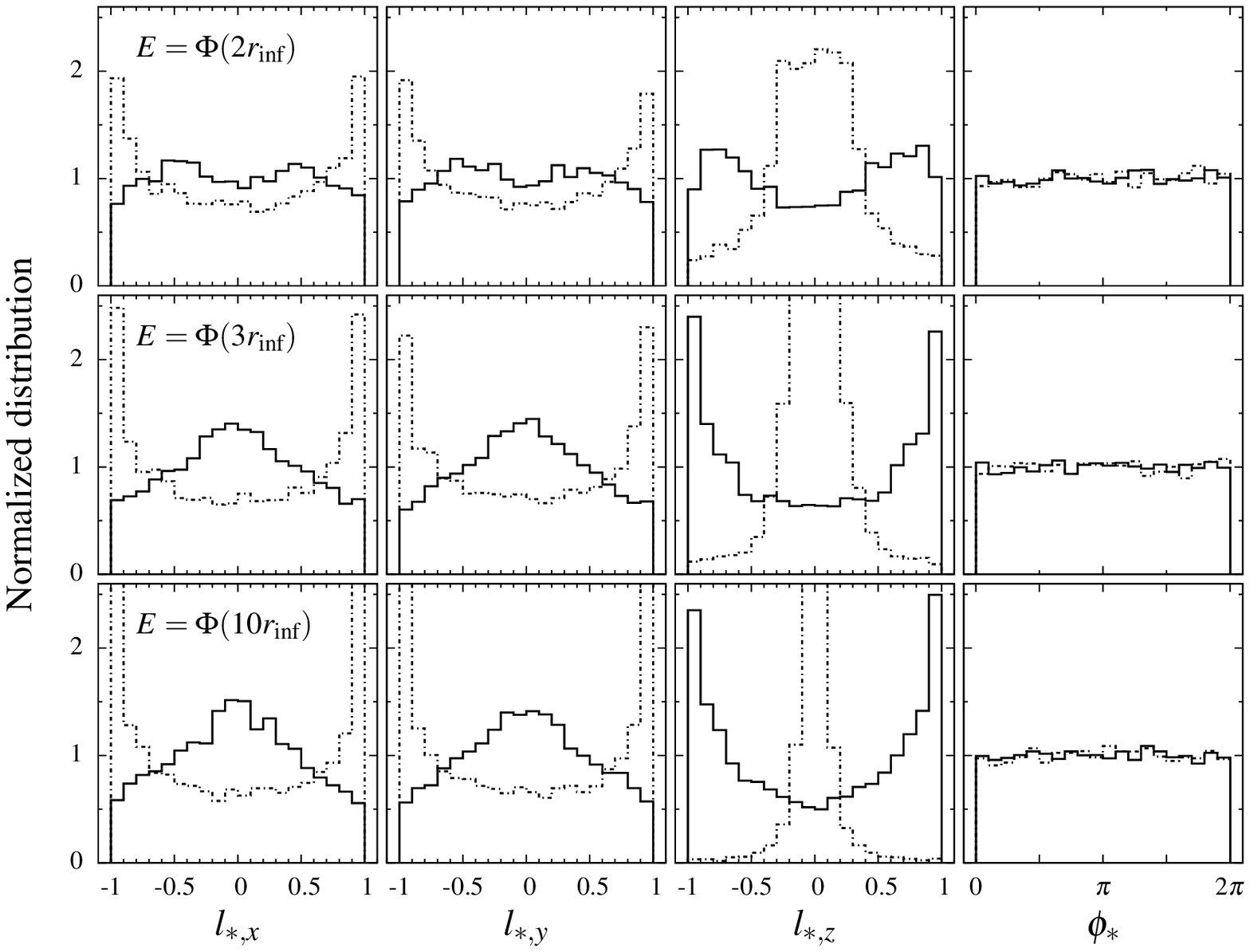}\centering
\caption{Histograms of the angular momentum orientation distribution of the
stars that can precess into the loss cone during the stellar precession stage
in axisymmetric galactic potential with $(\xi,\zeta)=(1,0.8)$. The lines have
the same meanings as those in Fig.~\ref{fig:precessiontriaxial}. 
As seen from the figure, in axisymmetric systems most of the stars have final
orbital orientations along the axisymmetric axis (i.e., $\pm z$-axis).
}
\label{fig:precessionaxisymmetric}
\end{figure}

Figure~\ref{fig:precessionaxisymmetric} shows an example for an axisymmetric
potential with $(\xi,\zeta)=(1,0.8)$, where the anisotropy of the distributions
does not decrease at high energy. As seen from the middle and the bottom
panels, the dotted line has a peak around $l_z=0$, as only the stars initially
located in the loss wedge ($|J_z|<J_{\rm lc}$) can precess into the loss cone
\citep{MT99}; and the solid line has peaks around $l_z=\pm 1$, indicating that
most of the stars have final orbital orientations along the axisymmetric axis
(i.e., $\pm z$-axis here).

\subsubsection{Rotational galaxies}

As mentioned above, the remnant of a galaxy merger is likely to be rotating
(e.g., see also \citealt{MM01}).
We generate the rotational property of a galaxy in the following way.  The
initial distribution of $l_{*,z}$ have different fractions for retrograde
orbits and prograde orbits, as done in \citet{GDS11}.  The probability that the
initial $\theta_*$ is randomly generated in the range of $[0,\pi/2]$ is denoted
by $P_+$, and then the probability in the range of $[\pi/2,\pi]$ is $1-P_+$.
For $P_+\ne 0.5$, the total angular momentum of the stars is non-zero and has a
residue along the $z$-axis. For a triaxial system, here the rotation axis is
assumed to align with one axis (i.e., $+z$-axis) of the system. 

\begin{figure}
\includegraphics[width=0.8\textwidth]{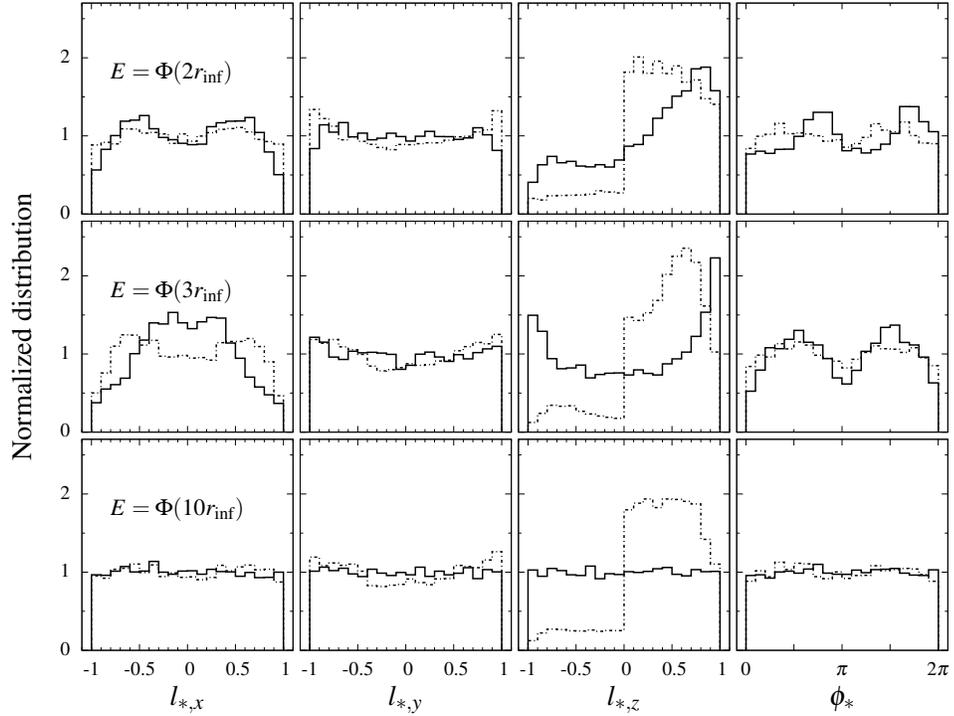}\centering
\caption{Histograms of the angular momentum orientation distribution of the
stars that can precess into the loss cone during the stellar precession stage
in triaxial galaxies with rotation $P_+=7/8$. The initial distributions (dotted
lines) are higher at $l_{*,z}>0$ than that at $l_{*,z}>0$, which indicates that
the system has a net rotation. As seen from the final distribution in the
bottom panel (solid line), the net rotation of high-energy stars is erased in
triaxial galaxies.
}
\label{fig:precessiontrirotation} \end{figure}

In spherical galaxies, the probability of $P_+$ keeps the same when the stars
approach to galactic centres, due to the conservation of angular momentum.
Figure~\ref{fig:precessiontrirotation} shows the result for triaxial galaxies,
where $P_+=7/8$ and the parameters $(R_{\rm c},\xi,\zeta)$ are the same as shown in
Figure~\ref{fig:precessiontriaxial}.  As seen from the figure, the initial
distributions of $l_{*,z}$ (the dot short-dashed curves) displays a strong
asymmetry of alignment, and the stars with $l_{*,z}>0$ outnumber those with
$l_{*,z}<0$ as the initial settings. As in Figure~\ref{fig:precessiontriaxial},
the final distributions also have a dependence on stellar energy.  When the
energy is low (top panel), the influence of the triaxial potential is mild, and
the final distributions follow the initial ones roughly. When the energy is
high (bottom panel), the alignment in the final distribution of $l_{*,z}$
decreases significantly; and the triaxial potential erases the alignment of the
initial stellar angular momenta because of stochastic orbits of the stars, for
which we call the {\it alignment-erasing effect} of the triaxiality here. 
Note that the degree of the galaxy net rotation adopted in this study is
probably stronger than that of realistic galaxies (e.g., \citealt{G11}); and
the alignment-erasing effect of the triaxiality, of course, should still hold
true for galaxies with smaller rotation.

\subsection{BBH orbital orientation evolution} \label{subsec:BBHevol}

We use the method described in Section~\ref{sec:method} to simulate the
evolution of the BBH orbital orientation in the three-body scattering stage.
For simplicity, we set the BBH eccentricity $e=0$ in the calculation.

We generate the initial orbital orientation of the BBH isotropically and trace
their evolution by interacting with a number of stars.  We show the BBH
orientation evolution as a function of the total mass of the interacting stars
(i.e., $Nm_*$), which is related to the BBH orbital decay by Equation
(\ref{eq:nscatter}).  We express the BBH orientation evolution through the
evolution of its angular momentum unit vector $(l_{\rmb,x},l_{\rmb,y},l_{\rmb,z})$, the
angles $\theta_\rmb\equiv \cos^{-1}{l_{\rmb,z}}$ and $\phi_\rmb$ is the azimuthal angle
of the vector projected onto the $x$-$y$ plane, and $\Delta\alpha$ is the angle
of the orientation deviated from its initial angular momentum direction. The
results are shown in Figures~\ref{fig:BBHevolspherical} and
\ref{fig:BBHevolnonspherical}.  In each figure, different curves display the
results for different initial BBH orientations.

\subsubsection{Spherical galaxies}

\begin{figure}
\begin{center}
\subfigure[]{\includegraphics[width=0.65\textwidth]{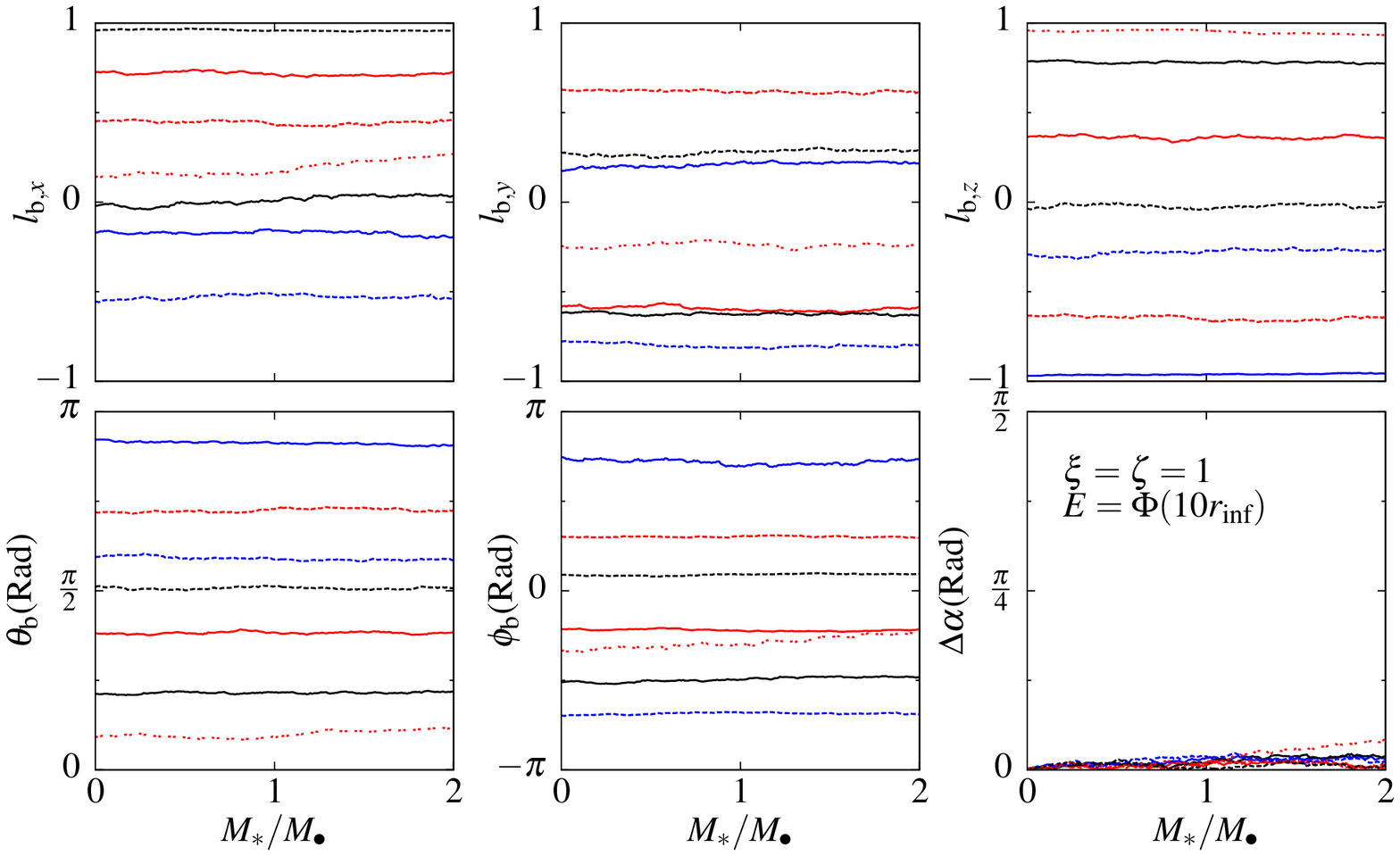}}
\subfigure[]{\includegraphics[width=0.65\textwidth]{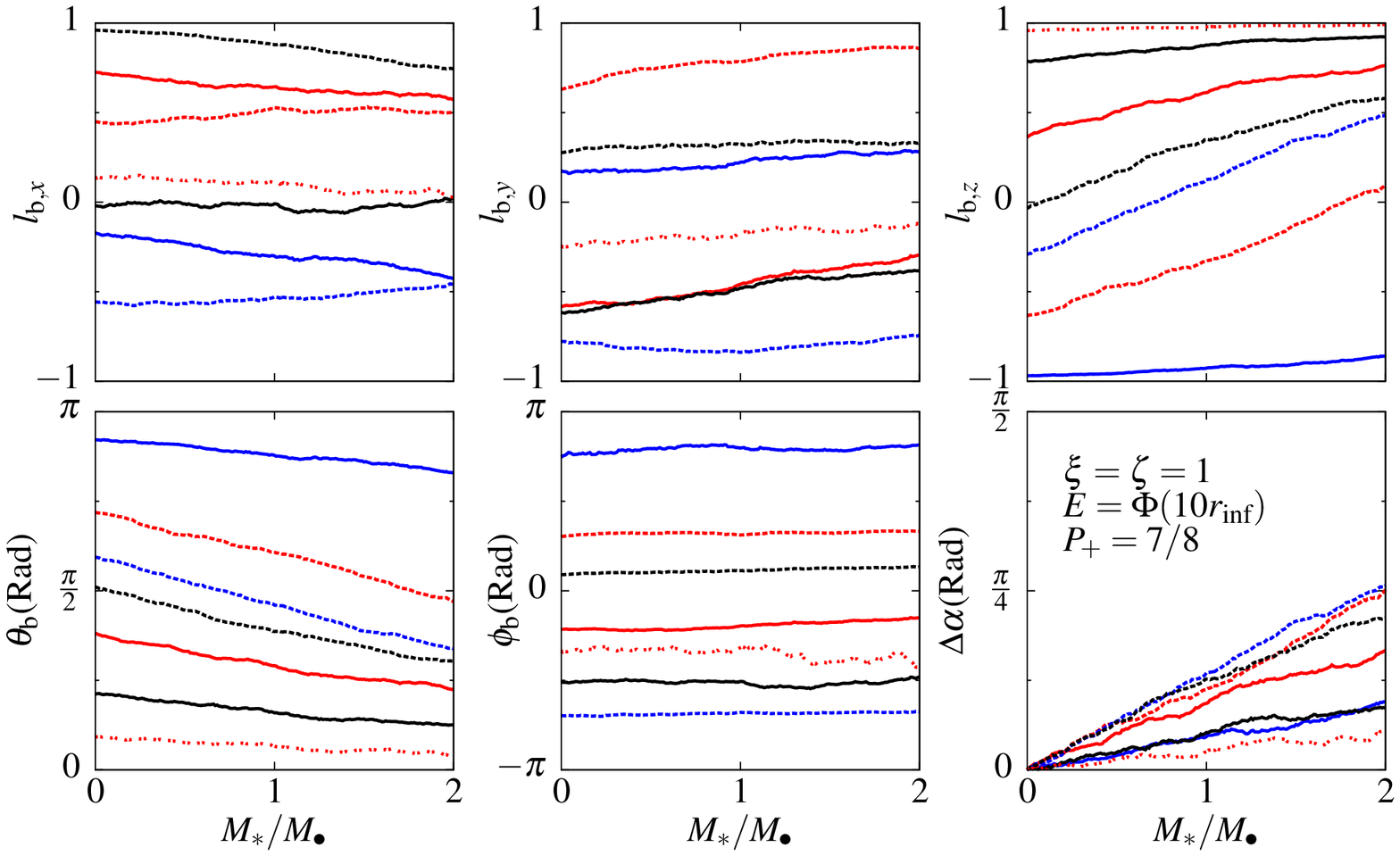}}
\end{center}
\caption{BBH orbital orientation evolution in spherical galaxies.
In the $x$-axis, $M_*=Nm_*$. Different curves represent different initial
orientations of the BBHs; and for visual clarity, different colors and
curve types are used to distinguish them. (a)
$P_+=1/2$, i.e., the total angular momentum of the stars in the system is zero;
(b) $P_+=7/8$, i.e., the system has a rotational property, with a non-zero
total stellar angular momentum along the $z$-axis. As mentioned in
Section~\ref{subsec:spherical}, statistically the results are not sensitive to
the value of $E$ if $E$ is high enough. Panel (b) indicates the alignment
effect of the BBH orbital orientations towards the rotational axis of the
spherical system, which is different from the random walk of the BBH
orientations shown in panel (a).
}
\label{fig:BBHevolspherical}
\end{figure}

Figure~\ref{fig:BBHevolspherical}(a) shows the BBH orbital orientation
evolution in spherical galaxies without rotation, and
Figure~\ref{fig:BBHevolspherical}(b) shows the results in those galaxies with
rotation $P_+=7/8$. As seen from Figure~\ref{fig:BBHevolspherical}(b),
$\theta_\rmb$ evolves toward 0, i.e., all of the orbital planes are reoriented
toward the direction of the total stellar angular momentum, which is consistent with the
result obtained by \citet{GDS11}.  The BBH orientation change $\Delta\alpha$ is
larger in panel (b) than that in panel (a).

\subsubsection{Non-spherical galaxies}

In Section~\ref{subsec:non-spherical}, we find that the kinematic distribution
of the stars that can come to the BBH vicinity can be anisotropic at the end of
the stellar precessing stage in non-spherical galaxies, and the anisotropy
depends on the stellar energy.

In triaxial galaxies, the distributions can be significantly anisotropic at
intermediate energies, but nearly isotropic at high energies. The value of the
energy where the anisotropy is significant depends on the non-spherical
properties of galaxies (e.g., triaxiality). In reality, the stars that can come
to the BBH vicinity have a distribution in energy, and the value of the energy
where the stars are the most numerously distributed depends mostly on the
radial distribution of the mass density of the galaxies, which may not be the
same as the energy where the anisotropy is significant. If the number of the
stars is dominated at an energy significantly higher or lower than the energy
where the anisotropy dominates, the BBH orientation evolution should undergo
random walks as that shown in spherical systems (e.g.,
Fig.~\ref{fig:BBHevolspherical}). In case that the number of the stars is
dominated at an energy where the anisotropy is significant, we get a chance to
see the effect of the anisotropy of the stellar kinematic distributions on the
BBH orbital orientation evolution.

\begin{figure} \begin{center}
\subfigure[]{\includegraphics[width=0.49\textwidth]{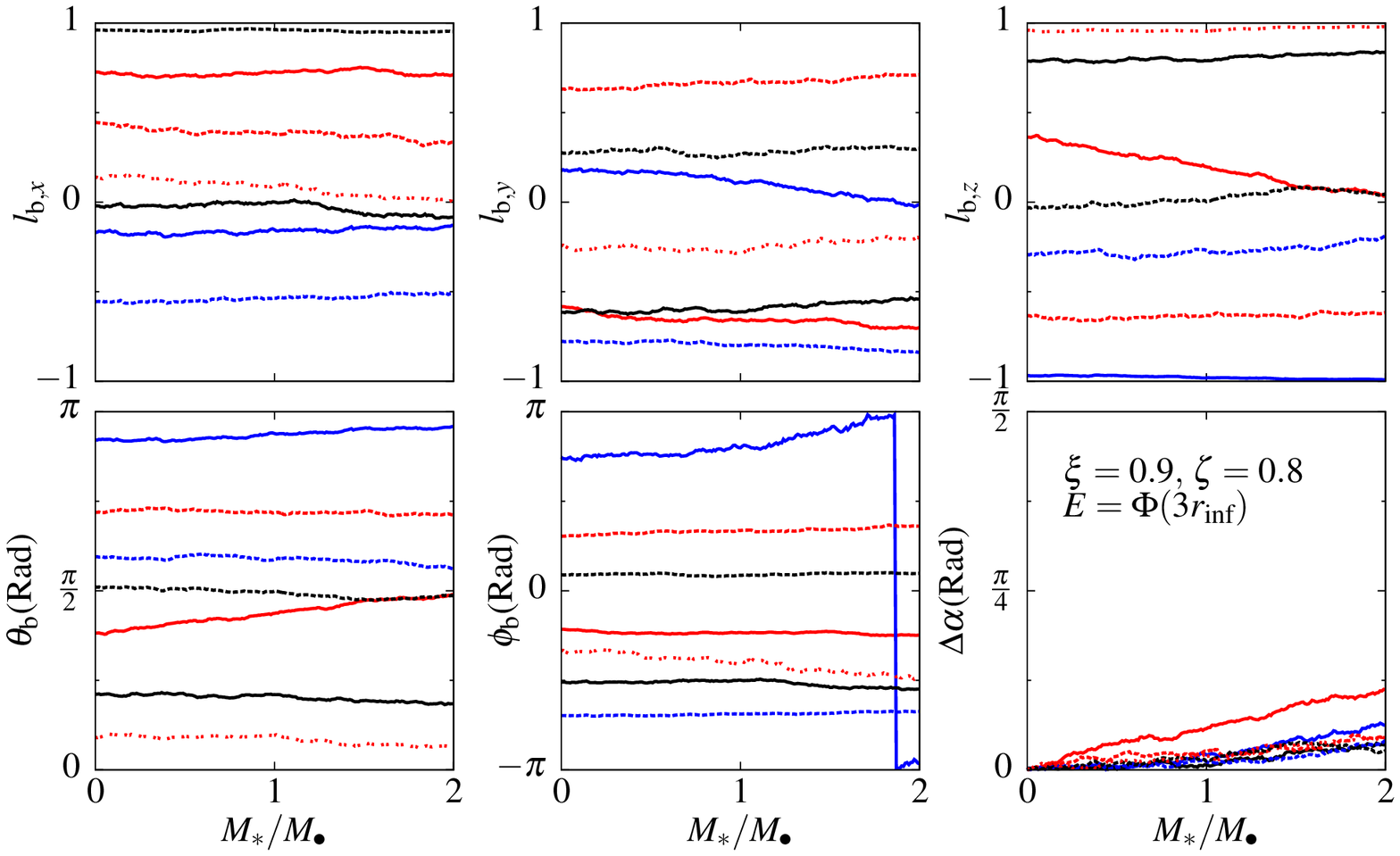}}
\subfigure[]{\includegraphics[width=0.49\textwidth]{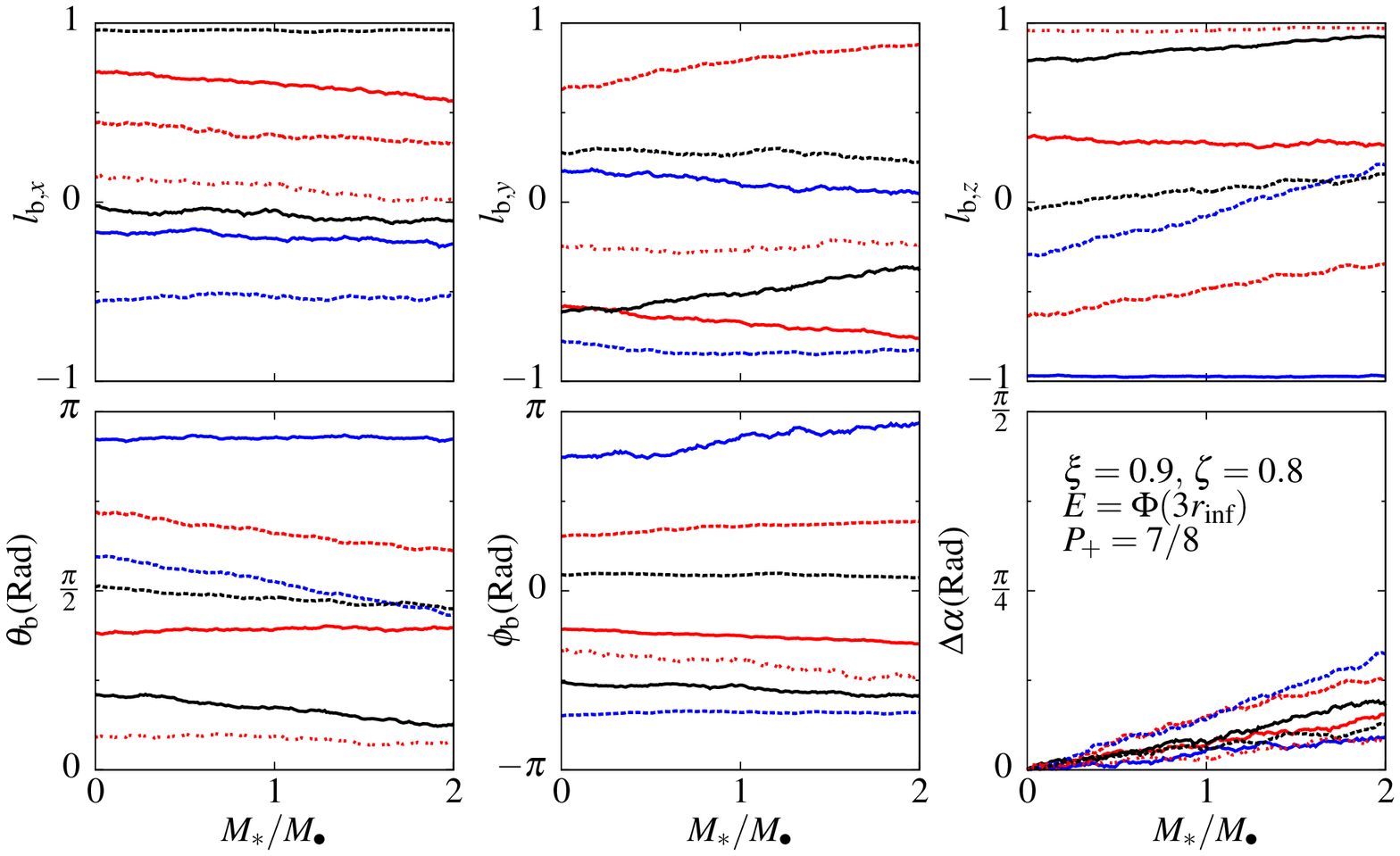}}
\subfigure[]{\includegraphics[width=0.49\textwidth]{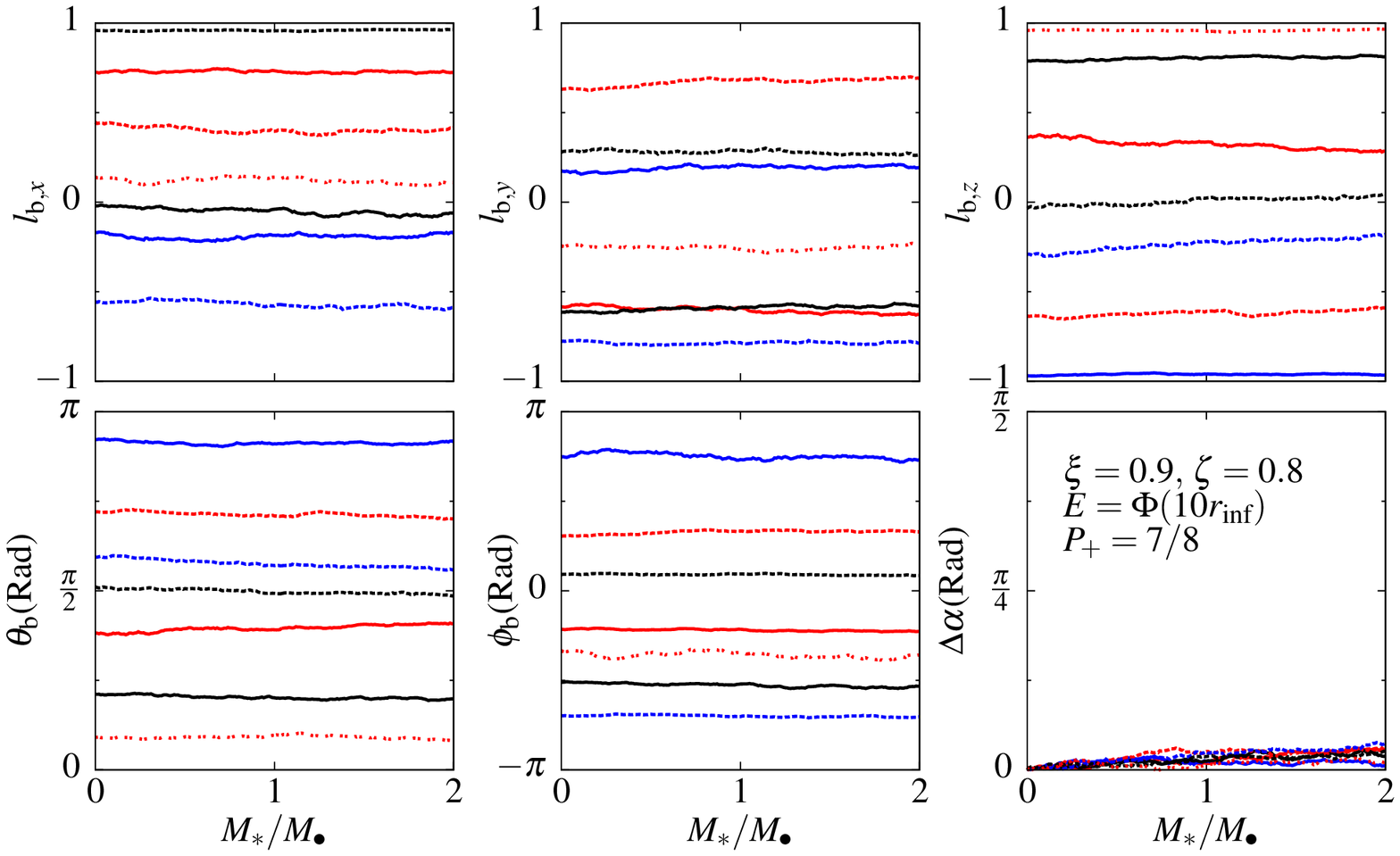}}
\subfigure[]{\includegraphics[width=0.49\textwidth]{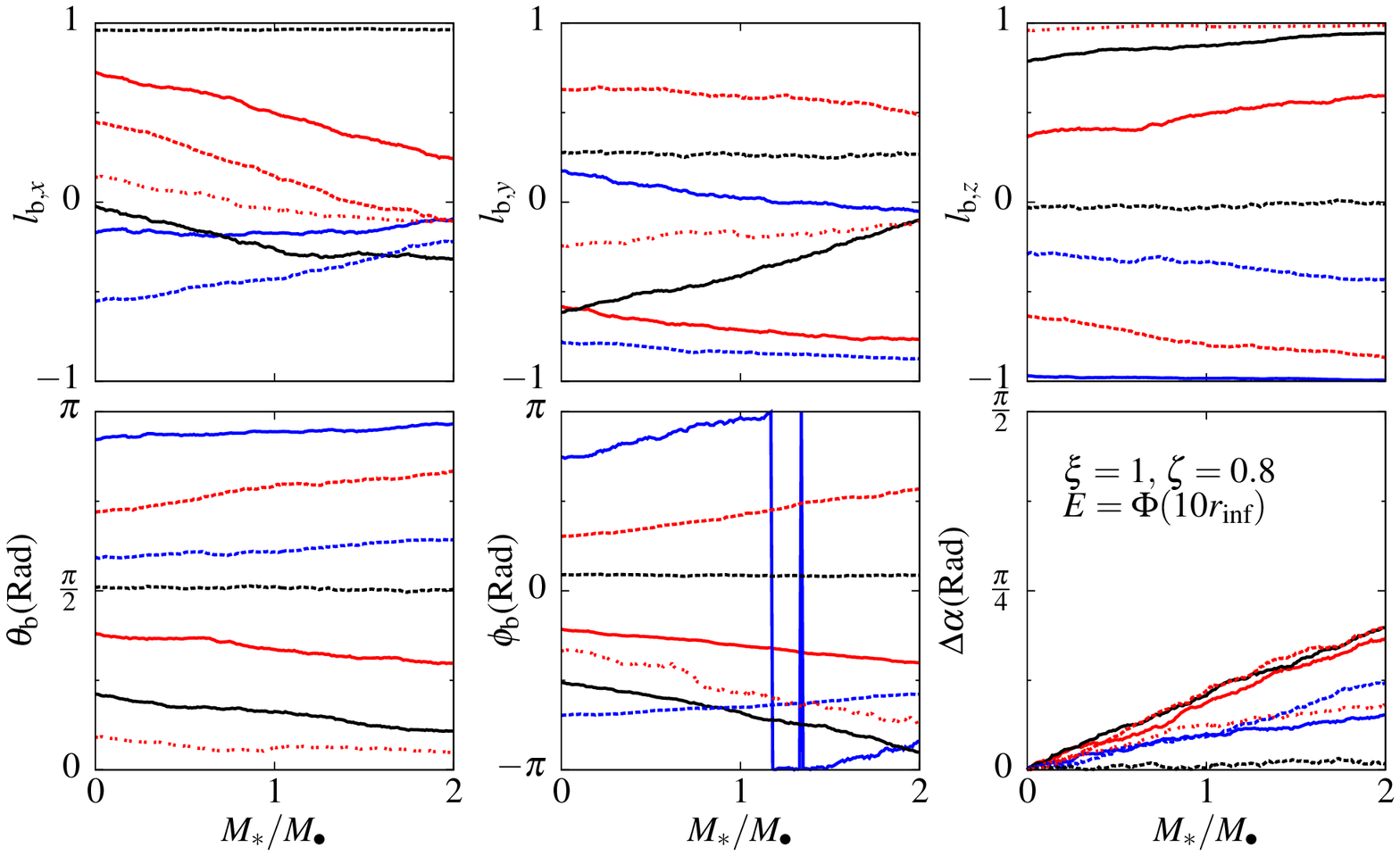}}
\end{center}
\caption{The BBH orientation evolution in non-spherical galaxies. The labels
and the curves have the same meanings as those
in Fig.~\ref{fig:BBHevolspherical}. (a) Triaxial
galaxies without rotation. The parameters are the same as those in the middle
panel of Fig.~\ref{fig:precessiontriaxial}.  (b) Triaxial galaxies with
rotation. The parameters are the same as those in the middle panel of
Fig.~\ref{fig:precessiontrirotation}. (c) Same as panel (b), but with different
stellar energy. The parameters are the same as those in the bottom panel of
Fig.~\ref{fig:precessiontrirotation}.  (d) Axisymmetric galaxies, with
parameters same as those in the bottom panel of
Fig.~\ref{fig:precessionaxisymmetric}. As seen from figure, the alignment
effect of the BBH orbital orientations towards the rotational axis can be
erased in triaxial systems (panel c), and the BBH orientations in
axisymmetric systems evolves towards alignment with the axisymmetric
axis (panel d).  
} \label{fig:BBHevolnonspherical} \end{figure}

Figure~\ref{fig:BBHevolnonspherical}(a)--(c) illustrates our results for
triaxial galaxies; and panels (b) and (c) are for those with rotating
properties, with different stellar energy, respectively. The initial stellar
kinematic distributions used for the three-body scattering stage correspond to
the final distributions shown in Figures~\ref{fig:precessiontriaxial} and
\ref{fig:precessiontrirotation}. 

As seen from Figure~\ref{fig:BBHevolnonspherical}(a), the orientation change of
the BBHs increases significantly, compared to the random walks in non-rotating
spherical systems shown in Figure~\ref{fig:BBHevolspherical}(a).  The total
angle change in Figure~\ref{fig:BBHevolnonspherical}(a) can be up to 0.4 rad
when a total $2 M_\bullet$ mass of stars are scattered; and after scattering
the stars with such a total mass, the BBH semimajor axis decays by a factor of
$\sim 500$ (see Eq.~\ref{eq:nscatter}).  The large angle change comes from the
relatively large change of $\theta_\rmb$ and $\phi_\rmb$, due to the anisotropy of
the initial kinematic distributions of the interacting stars (see the middle
panel of Fig.~\ref{fig:precessiontriaxial}). The effect is more clearly shown
for the axisymmetric case in Figure~\ref{fig:BBHevolnonspherical}(d) below.

The BBH orientation changes for rotating triaxial galaxies shown in
Figures~\ref{fig:BBHevolnonspherical}(b)-(c) are smaller than those for
rotating spherical systems shown in Figure~\ref{fig:BBHevolspherical}(b).  The
reduced change originates from the alignment-erasing effect of the triaxial
potential on the kinematic distributions of the interacting stars, as mentioned
in Section~\ref{subsec:non-spherical} (see
Fig.~\ref{fig:precessiontrirotation}).  The effect is stronger for the case
with high stellar energy shown in Figure~\ref{fig:BBHevolnonspherical}(c),
where the final stellar kinematic distribution obtained at the stellar
precessing stage is close to isotropic and the magnitude of the BBH orientation
evolution is close to the random walk results shown in
Figure~\ref{fig:BBHevolspherical}(a). Although the alignment-erasing effect
shown in Figure~\ref{fig:BBHevolnonspherical} is done for the BBH mass ratio
$q=0.1$, it is plausible to expect that it also exist for some other $q$, which
is supported by our numerical tests (e.g., for $q=0.3$).

Figure~\ref{fig:BBHevolnonspherical}(d) shows the BBH orientation evolution in
axisymmetric galaxies, which is much higher than that for non-rotating
spherical galaxies shown in Figure~\ref{fig:BBHevolspherical}(a). As seen from
Figure~\ref{fig:BBHevolnonspherical}(d), the orientation change comes from the
two following parts: (1) the motion of the orientation towards the axisymmetric
axis ($\pm z$-axis); that is, $\theta_\rmb$ evolves towards $0$ if initially
$0\la\theta_\rmb\la\pi/2$ or towards $\pi$ if initially $\pi/2\la\theta_\rmb\la\pi$;
and (2) the retrograde precession in $\phi_\rmb$ around the axisymmetric axis;
that is, $\phi_\rmb$ statistically tend to increase for $\pi/2\la\theta_\rmb\la\pi$
and decrease for $0\la\theta_\rmb\la\pi/2$. As seen from
Figure~\ref{fig:precessionaxisymmetric}, the stars passing by the vicinity of
the BBH in axisymmetric potential are inclined to distribute in the plane
perpendicular to the axisymmetric axis.  The motion of the BBH orientation
towards the axisymmetric axis ($\pm z$-axis) suggests that the average BBH
orientation alignment magnitude towards the orbital angular momentum of a star
caused by interacting with the star is statistically larger if the star is on a
prograde orbit (with relative inclination to the BBH orbit, denoted by $\beta$
here; $0<\beta<\pi/2$) than the alignment magnitude if the star is on a
retrograde orbit (with relative inclination $\pi-\beta$).  The retrograde
precession of $\phi_\rmb$ can be explained through the gravitational potential of
the planar mass distribution of the passing-by stars, with dynamics analogical
to the precession of a spinning top under the torque induced by the earth's
gravity. If the stars in the axisymmetric galaxies have a net rotation with
non-zero total angular momentum along the axisymmetric axis, the BBH orbital
orientation is expected to evolve towards the direction of the total angular
momentum as shown in Figure~\ref{fig:BBHevolspherical}(b); and for simplicity,
we do not show the calculation results for this case here. 

\subsection{BBH orbital orientation distributions} \label{subsec:BBHdistr}

\begin{figure} 
\includegraphics[width=0.6\textwidth]{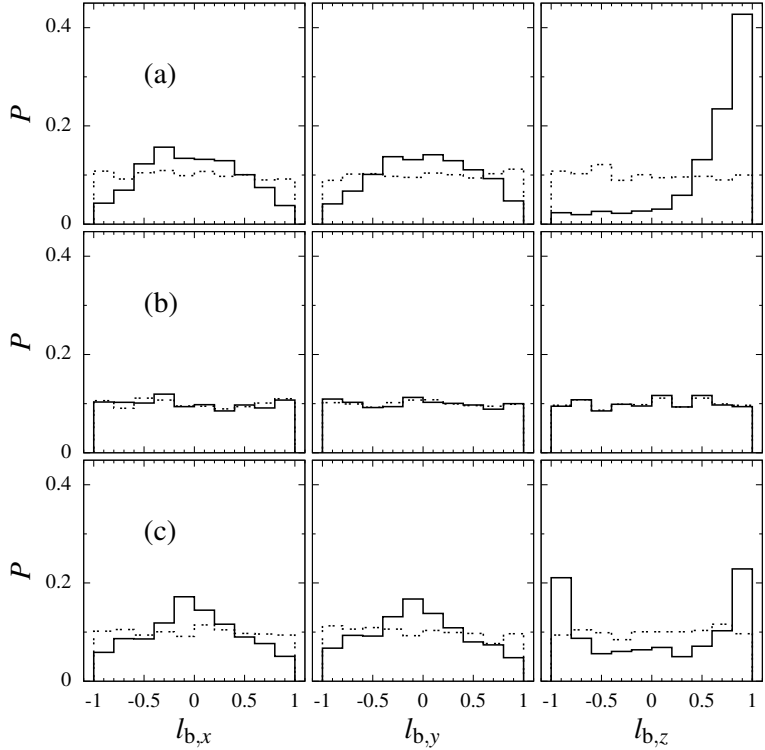}\centering
\caption{The simulated BBH orbital orientation distributions.  Different rows
are for different systems: (a) rotating spherical systems; (b) rotating
triaxial galaxies with $(\xi,\zeta)=(0.9,0.8)$ and $E=\Phi(10r_{\rm inf})$; and (c)
axisymmetric galaxies with $(\xi,\zeta)=(1,0.8)$.  The dotted lines represent
the initial isotropic distributions of the BBH orientations, and the solid lines
in panels (a)--(c) represent the distributions obtained at the end of
simulation shown in Figs.~\ref{fig:BBHevolspherical}(b), and
\ref{fig:BBHevolnonspherical}(c)--(d), respectively. The number of the
simulated BBHs for each panel is 1500. The distributions of the solid lines
illustrate the alignment of the BBH orientations along with the rotation axis
in rotating spherical systems (panel a), the alignment-erasing effect in triaxial
systems (panel b), and the alignment along with the axisymmetric axis in
axisymmetric systems (panel c).
}
\label{fig:BBHdis}
\end{figure}

To see the effects of different potentials on the BBH orbital orientation
distributions, we generate a sample of 1500 BBHs and obtain their orientation
distributions after they interact with a large number of stars with total mass
$\sim2$ times the BBH mass. The initial orientations of the BBHs are assumed to
distribute isotropically, for a large range of initial conditions instead
of limiting to some specific directions.  Figure~\ref{fig:BBHdis} shows our
results of the BBH orbital orientation distributions in different systems.
Panel (a) is for rotating spherical systems, where the distribution of $l_z$
peaks at $l_z\simeq 1$, i.e., the reorientation effect is strong and most of
the BBHs orient toward the rotating axis.  Panel (b) shows the results for
rotating triaxial galaxies, where the BBH orientations are isotropically
distributed due to the alignment-erasing effect.  Panel (c) is for the
axisymmetric galaxies, where the distribution peaks at $l_{\rmb,z}=\pm 1$ and most
of the BBH orientations are aligned along the axisymmetric axis ($\pm z$-axis).

In triaxial galaxies, for the general case in which the kinematic distribution
of the stars that can pass by the vicinity of the BBH is isotropic at the end
of the stellar precessing stage (e.g., bottom panel of
Fig.~\ref{fig:precessiontriaxial}), the magnitude of the BBH orientation
evolution is small as shown in Figure~\ref{fig:BBHevolnonspherical}(c).  Thus,
the isotropic distribution of the BBH orientations shown in
Figure~\ref{fig:BBHdis}(b) is due to the initial isotropy of the BBH
orientations set in the calculation. If the BBH initial orientation
distribution is anisotropic, the anisotropy of the distribution maintains as
the orientation evolution has a small magnitude.

In axisymmetric galaxies, the BBH orientation distribution shown in
Fig.~\ref{fig:BBHdis}(c) is not affected by the precession in $\phi_\rmb$,
although the total orientation angle change $\Delta\alpha$ may be different by
different precession rates.

\subsection{Discussion} 
\label{subsec:discussion}

Below we discuss the results and some assumptions made in our method.

In our calculations above, we set a fixed stellar mass with
$m_*=10^{-4}M_\bullet$ for simplicity.  If the BBH orientation evolution is
like random walks (e.g., Figs.~\ref{fig:BBHevolspherical}(a) and
\ref{fig:BBHevolnonspherical}(c)), the magnitude of the orientation change is
expected to be proportional to $\sqrt{N\langle\delta\alpha^2\rangle}\propto
\sqrt{m_*/M_\bullet}$, as indicated in Equation (\ref{eq:delta_theta}), where
$N\propto M_\bullet/m_*$ and ${\langle\delta\alpha^2\rangle}^{1/2}\propto
m_*/M_\bullet$ are used.  For the orientation evolution in rotating spherical
systems (e.g., Fig.~\ref{fig:BBHevolspherical}(b)), the magnitude of the
orientation change is expected to be insensitive to the value of $m_*$, as the
BBH orbital alignment towards the rotational axis is approximately proportional
to $N$ instead of $\sqrt{N}$ (see also \citealt{GDS11}). For other cases (e.g.,
Fig.~\ref{fig:BBHevolnonspherical}(a), (b), and (d)), our numerical tests also
support that the magnitude of the orientation change is insensitive to the
value of $m_*$ for $m_*/M_\bullet$ ranging from $10^{-6}$ to $10^{-4}$.

In the cases that the BBH orientation evolves like random walks (e.g.,
Figs.~\ref{fig:BBHevolspherical}(a) and \ref{fig:BBHevolnonspherical}(c)), the
dependence of the orientation change on the BBH mass ratio $q$ and eccentricity
$e$ is indicated in Equation (\ref{eq:delta_theta}). For the other cases, the
quantitative dependence on $q$ and $e$ is not obvious, but it can be achieved
by performing numerical simulations and qualitative analysis. 
The alignment-erasing effect discussed for triaxial systems above should still
exist even for different $q$ and $e$.  We expect that for sufficiently small
$q$ and high $e$, the random walks of the BBH orientation evolution may surpass
its alignment effect along the rotational axis even in spherical or
axisymmetric galaxies.

Note that the net rotation property of a stellar system discussed here is
different from the global pattern rotation of a galaxy shape. In the case that
the triaxial galactic potential (e.g., Eq.~\ref{eq:tripotential}) has a global
pattern rotation around one of its axis (e.g., \citealt{DVM11}), we expect that
the effects on the the stellar kinematic distributions discussed in this study
would be qualitatively close to the results obtained by approximating the
rotating triaxial potential as a potential axisymmetric around the pattern
rotation axis if the speed is fast enough, while the effects are little if the
pattern speed is low.

\begin{figure}\begin{center}
\includegraphics[width=0.6\textwidth]{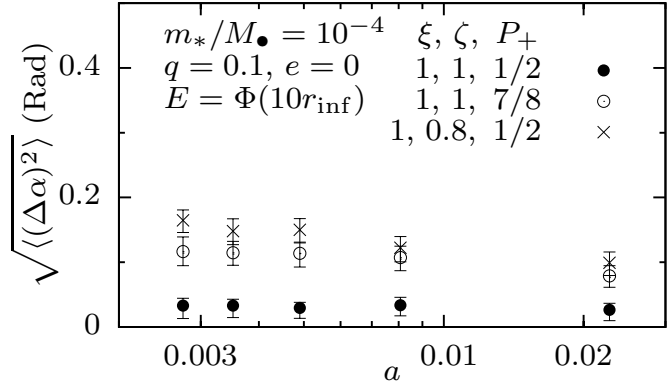}
\end{center}
\caption{The rms of the simulated orientation change of hard BBHs obtained in
one hardening time. The points are our simulation results at the different
fixed BBH semimajor axis.  The open and the solid circles are for spherical
systems ($\xi=1,\zeta=1$) with rotation ($P_+=7/8$) and without rotation
($P_+=1/2$), respectively; and the crosses are for axisymmetric systems without
rotation.  Each point is the rms of orientation change of 100 hard BBHs. In the
spherical systems, the initial orientations of the BBHs are generated randomly;
and in the axisymmetric systems, we fix their values of $\theta_\rmb$($=\pi/4$) to
see the effects solely due to the change of the semimajor axes. The error-bar
of each point has the same meaning as those in
Fig.~\ref{fig:deltathetaspherical}.  As seen from this figure, the rms of the
orientation change of hard BBHs is not affected much by different $a$.  See
more in Section~\ref{subsec:discussion}.  }
\label{fig:afixing} \end{figure}

In our method, the semimajor axis and the eccentricity of the BBH is fixed
during the three-body scattering stage. In reality, the BBH orbit shrinks due
to interactions with stars. By using the similar method above, we calculate the
BBH orientation evolution at different fixed BBH semi-major axes.  We find that
the orientation changes at different $a$ do not differ significantly in
spherical systems (see circles in Fig.~\ref{fig:afixing}), which is consistent
with Equation (\ref{eq:delta_theta}) that is not sensitive to the parameter
$a$.  Figure~\ref{fig:afixing} shows that the orientation changes at different
$a$ are also mild in axisymmetric systems (see crosses in the figure); and the
changes result mainly from different precession speeds in $\phi_\rmb$, which do
not affect the BBH orientation distribution (e.g., as shown in
~\ref{fig:BBHdis}c).  Regarding the BBH orientation evolution, the cases
illustrated in Figure~\ref{fig:afixing} are the extreme ones among the cases
shown in Figures~\ref{fig:BBHevolspherical}--\ref{fig:BBHevolnonspherical},
which suggests that our main conclusions should not be affected by the fixing
of $a$. Regarding the BBH eccentricity evolution, it is plausible to ignore it
in an isotropic stellar system if the initial BBH eccentricity is low, as
\citet{Q96} shows that the change of the BBH eccentricity in such a case is not
significant (see also the simulation results of different initial settings in
\citealt{A10,H02}).
In kinematically anisotropic
spherical systems with stars mostly counter-rotating with the BBH orbit, some
simulations show that the BBH eccentricity could increase significantly
\citep{SGD11,A10}; and we expect that the significant increase of the BBH
eccentricity can be decreased or erased if the systems are triaxial, by
applying the same reason for the alignment-erasing effect obtained in this
paper.

As mentioned in Section~\ref{sec:introduction}, before a BBH becomes bound and
hard, the two BHs sink into the galactic centre through dynamical friction, and
their orbital orientation is likely to be changed in a non-spherical galaxy
during the dynamical friction stage.  We construct a simple model to simulate
the inspiraling of a relatively small BH from a large scale (e.g., $10^4R_{\rm c}$)
into the galactic centre during the dynamical friction stage, as modeled in
Section 4 in \citet{yu02}. We assume that the distribution of the orbital
orientations of the small BHs is isotropic initially at the large scale, and we
find that in triaxial galaxies (e.g., with triaxiality parameters as those in
Fig.~\ref{fig:BBHdis}(b)) the BBH orbital orientation distribution is still
isotropic when the two BHs becomes bound, as assumed for the initial BBH
orientation distributions in Figure~\ref{fig:BBHdis}.  In significantly
flattened axisymmetric galaxies (e.g., $\xi=1$, $\zeta=0.7$), we find that the
distribution tends to cluster toward the axisymmetric axis; and the BBH
orientation distribution obtained from Figure~\ref{fig:BBHdis}(c) are not
affected qualitatively, unless the shape of the galaxy changes with the
galactic radius significantly and the axisymmetric axis of the galactic centre
is significantly different from that of the galaxy at the large scale.

As mentioned in Section~\ref{sec:method}, the Newtonian mechanics is used in
our calculations of the motion of the BBH and the stars. To see how the general
relativistic effect could affect the calculation results in the three-body
scattering stage, we estimate the potential contribution from the
post-Newtonian terms by orders of magnitude as follows.  Consider a star that
starts with a low velocity passes by the BBH at a distance $r\simeq a$. The
energy and the angular momentum change of the star comes mainly from the
interaction with the smaller BH if $q\ll 1$.  First, with only using the
Newtonian mechanics, the specific interaction force on the star $F\sim
Gm_2/a^2$ acts for a time $\delta t\sim (a^3/GM_\bullet)^{1/2}$ to produce a
velocity change $\delta v\sim F\delta t$ and thus result in an orientation
change of the BBH $\langle \delta\alpha^2 \rangle ^{1/2}\propto (v\delta
v)^{1/2}/J_\rmb$, where $v\sim (2GM_\bullet/a)^{1/2}$ and $J_\rmb$ is the
orbital angular momentum of the BBH. Then, if considering the contribution from
the post-Newtonian terms, the relative change in $\langle \delta\alpha^2
\rangle^{1/2}$ is $\sim \frac{1}{2}[\frac{\Delta v}{v}+\frac{\Delta(\delta
v)}{\delta v}-2\frac{\Delta J_\rmb}{J_\rmb}]$, where $\Delta v$,
$\Delta(\delta v)$, and $\Delta J_\rmb$ are the changes in $v$, $\delta v$,
and $J_\rmb$ due to the post-Newtonian terms, respectively.  To approximate
the contribution from the post-Newtonian terms, we replace the gravitational
potential $-\frac{Gm}{r}$ of a particle with mass $m$ by $-\frac{Gm}{r-r_{\rm g}}$,
where $r_{\rm g}=2Gm/c^2$ and $c$ is the speed of light \citep{BW80}. Thus we have
$\frac{\delta v}{v}\sim \frac{GM_\bullet}{a c^2}$, $\frac{\Delta(\delta
v)}{\delta v}\sim -\frac{2GM_\bullet}{ac^2}$, and $\frac{\Delta
J_\rmb}{J_\rmb}\sim \frac{2GM_\bullet}{ac^2}$; and the absolute value of the
relative change in $\langle \delta\alpha^2 \rangle^{1/2}$ is $\sim
\frac{5}{2}\frac{GM_\bullet}{ac^2}\sim 4.4\times
10^{-6}q^{-1}(\sigmac/200\kms)^2(a/a_{\rm h})^{-1}$, where the definition of $a_{\rm h}$
in Equation (\ref{eq:ah}) is used. So the effect of the post-Newtonian terms is
generally small enough to be negligible for the study of this paper.

Gas-poor mergers can occupy a significant fraction of galaxy mergers (e.g.,
\citealt{B12,L08}). However, a significant fraction of galaxy mergers can also be
gas-rich.  In a merger with sufficient gas, it is expected that the orbital
orientation of the BBH always evolve towards co-alignment with the angular
momentum direction of its circumbinary disk. However, the role of the gas
depends on the amount of gas available. On a timescale short compared with the
mass growth timescale of the BBH, the BBH orientation may maintain the
counter-alignment if the BBH is initially sufficiently retrograde with the disk
\citep{N11}; and each BH may be surrounded by its own accretion disk (e.g.,
\citealt{YL01,D07,HMH08}), which does not necessarily co-align with the
circumbinary disk.

\section{Summary} \label{sec:summary}

In this paper, we study the orbital orientation evolution of BBH systems, and
mainly focus on their evolution in triaxial (and axisymmetric) systems. For
spherical stellar systems, we have reproduced the result that the orientation
of a BBH undergoes random walks in kinematically isotropic systems. And in
rotating spherical galaxies where the stars have a non-zero total angular
momentum, the BBH orientation change is larger and its orbital direction
reorients toward the direction of that total angular momentum.  In triaxial
systems, the initial angular momenta of the stars that can precess to the very
vicinity of a BBH can be far larger than those of the stars initially in the
loss cone.  The directions of the angular momenta of the stars when they
precess to the vicinity of a BBH can be anisotropic due to the torque induced
by the triaxial potential, and the degree of the anisotropy depends on the
stellar energy as well as the triaxiality or the flattening of the galactic
gravitational potential.  In axisymmetric galaxies, the angular momenta of the
stars are inclined to align to the axisymmetric axis of the system. If the
anisotropy of the kinematic distributions of the interacting stars is
significant, the orientation evolution of a BBH is different from random walks
obtained in non-rotating spherical galaxies. In axisymmetric galaxies, the
evolution of the BBH orientation includes the alignment towards the
axisymmetric axis and the precession around the axisymmetric axis in a
direction retrograde to the BBH orbit. 

We find that the triaxial potential has an {\it alignment-erasing} effect, if
most of the stars that can come to the vicinity of the BBH have significantly
high energy. In this case, the kinematic distribution of the stars when they
come to the BBH vicinity is close to isotropic even if they are initially
anisotropic (e.g., in a rotating system). Due to the alignment-erasing effect,
the alignment of the BBH orientations towards the rotation axis of a system can
be decreased significantly in triaxial galaxies.

If orbital orientation distributions of BBHs are isotropic when they become
hard, their distributions may maintain isotropic after interacting with
sufficiently numerous stars in significantly triaxial systems; however, they
can be anisotropic in axisymmetric systems, and the number of BBHs with angular
momentum directions near the axisymmetric axis is enhanced. If a BBH with
comparable mass components (e.g., formed in a major merger of two galaxies)
merges, the direction of the BBH orbital angular momentum is likely to dominate
the spin direction of the merged BH remnant, and also dominate the jet
direction if some gas sink to the vicinity of the merged BH and a jet is
produced.  If the galaxy is significantly triaxial, the alignment-erasing
effect obtained in this paper would imply that there is no preferred
orientation of host galaxies and radio jets, but a preferential alignment in
relatively axisymmetric galaxies if the axisymmetric axis of the system does
not change significantly with galactic radii. 

Regarding the tendency for the axis of the radio emission to align with an axis
of the galaxy starlight, a dependence on the galaxy properties was possibly
revealed in observations. Such a tendency was found in weak radio-loud AGNs,
but not among the radio-louder objects possibly hosted in triaxial elliptical
galaxies (e.g., \citealt{BB10,SS09}).

This preliminary study mainly uses simple hybrid models for the galactic
gravitational potential, stellar distribution, and rotation of the system to
illustrate the related effects; and the simplification of the model should not
a concern for the purpose of this paper, although a more sophistical
self-consistent model of these properties needs to be used in realistic
galaxies.

We thank Youjun Lu for helpful discussion and comments. This research was
supported in part by the National Natural Science Foundation of China under
nos.\ 10973001, 11273004.

\end{document}